\documentclass[sigconf, nonacm]{acmart}

\newcommand{\sun}[1]{{\textcolor{black}{ #1}}}
\newcommand{\pp}[1]{{\textcolor{black}{ #1}}}

\newcommand\vldbdoi{XX.XX/XXX.XX}
\newcommand\vldbpages{XXX-XXX}
\newcommand\vldbvolume{16}
\newcommand\vldbissue{5}
\newcommand\vldbyear{2023}
\newcommand\vldbauthors{\authors}
\newcommand\vldbtitle{\shorttitle} 
\newcommand\vldbavailabilityurl{}
\newcommand\vldbpagestyle{empty} 

\usepackage[ruled,linesnumbered]{algorithm2e}

\usepackage{pifont}

\begin{document}

\title{Async-fork: Mitigating Query Latency Spikes Incurred by the Fork-based Snapshot Mechanism from the OS Level}

%%
%% The "author" command and its associated commands are used to define the authors and their affiliations.
%\author{Ben Trovato}
%\affiliation{%
%  \institution{Institute for Clarity in Documentation}
%  \streetaddress{P.O. Box 1212}
%  \city{Dublin}
%  \state{Ireland}
%  \postcode{43017-6221}
%}
%\email{trovato@corporation.com}

\author{$^*$Pu Pang$^{1,2}$, $^*$Gang Deng$^2$, Kaihao Bai$^{1,2}$, Quan Chen$^1$, Shixuan Sun$^3$, Bo Liu$^1$, Yu Xu$^2$, Hongbo Yao$^2$, Zhengheng Wang$^2$, Xiyu Wang$^2$, Zheng Liu$^2$, Zhuo Song$^{1,2}$
, Yong Yang$^2$, Tao Ma$^2$, Minyi Guo$^1$}
\affiliation{$^1$Department of Computer Science and Engineering, Shanghai Jiao Tong University, China}
\affiliation{$^2$Alibaba Group, China}
\affiliation{$^3$National University of Singapore, Singapore}
\email{{avengerispp, asbaikaihao, chen-quan, boliu98, myguo}@sjtu.edu.cn, sunsx@comp.nus.edu.sg}
\email{{denggang.dg, qiyu.xy, yuanzhi.yhb, zhengheng.wzh, xiyu.wxy, wenqing.lz, songzhuo.sz, boyu.mt}@alibaba-inc.com}

%%
%% The abstract is a short summary of the work to be presented in the
%% article.
\begin{abstract}
In-memory key-value stores (IMKVSes) serve many online applications because of their efficiency. To support data backup, popular industrial IMKVSes periodically take a point-in-time snapshot of the in-memory data with the system call \emph{fork}. However, this mechanism can result in latency spikes for queries arriving during the snapshot period because \emph{fork} leads the engine into the kernel mode in which the engine is out-of-service for queries. In contrast to existing research focusing on optimizing snapshot algorithms, we optimize the fork operation to address the latency spikes problem from the operating system (OS) level, while keeping the data persistent mechanism in IMKVSes unchanged. Specifically, we first conduct an in-depth study to reveal the impact of the fork operation as well as the optimization techniques on query latency. Based on findings in the study, we propose Async-fork to offload the work of copying the page table from the engine (the parent process) to the child process as copying the page table dominates the execution time of \emph{fork}. To keep data consistent between the parent and the child, we design the proactive synchronization strategy. Async-fork is implemented in the Linux kernel and deployed into the online Redis database in public clouds. Our experiment results show that compared with the default fork method in OS, Async-fork reduces the tail latency of queries arriving during the snapshot period by 81.76\% on an 8GB instance and 99.84\% on a 64GB instance.

\end{abstract}

\maketitle

%%% do not modify the following VLDB block %%
%%% VLDB block start %%%
\pagestyle{\vldbpagestyle}
\begingroup\small\noindent\raggedright\textbf{PVLDB Reference Format:}\\
\vldbauthors. \vldbtitle. PVLDB, \vldbvolume(\vldbissue): \vldbpages, \vldbyear.\\
\href{https://doi.org/\vldbdoi}{doi:\vldbdoi}
\endgroup
\begingroup
\renewcommand\thefootnote{}\footnote{\noindent 
$^*$These authors contributed equally to this work. \\
This work is licensed under the Creative Commons BY-NC-ND 4.0 International License. Visit \url{https://creativecommons.org/licenses/by-nc-nd/4.0/} to view a copy of this license. For any use beyond those covered by this license, obtain permission by emailing \href{mailto:info@vldb.org}{info@vldb.org}. Copyright is held by the owner/author(s). Publication rights licensed to the VLDB Endowment. \\
\raggedright Proceedings of the VLDB Endowment, Vol. \vldbvolume, No. \vldbissue\ %
ISSN 2150-8097. \\
\href{https://doi.org/\vldbdoi}{doi:\vldbdoi} \\
}\addtocounter{footnote}{-1}\endgroup
%%% VLDB block end %%%

%%% do not modify the following VLDB block %%
%%% VLDB block start %%%
%\ifdefempty{\vldbavailabilityurl}{}{
%\vspace{.3cm}
%\begingroup\small\noindent\raggedright\textbf{PVLDB Artifact Availability:}\\
%The source code, data, and/or other artifacts have been made available at %\url{\vldbavailabilityurl}.
%\endgroup
%}
%%% VLDB block end %%%

\section{Introduction}

In-memory key-value stores (IMKVSes) are widely used in real-world applications, especially online services (e.g., e-commerce and social network), because of their ultra-fast query processing speed. For example, Memcached~\cite{memcached}, Redis~\cite{redis}, KeyDB~\cite{keydb} and their variants~\cite{chandramouli2018faster, fan2013memc3, harris2010distributed, bailey2013exploring} have been deployed in production environments of big internet companies such as Facebook, Amazon and Twitter. As all data resides in memory, the data persistent function is a key feature of IMKVSes for data backup.

\begin{figure*}[!t]
  \centering
  \includegraphics[width=2.0\columnwidth]{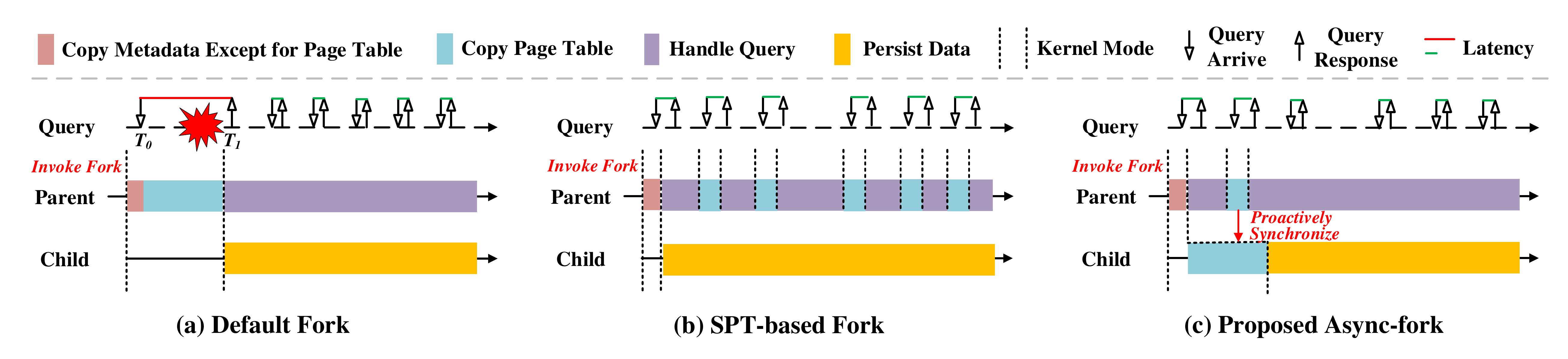}
%\vspace{-2mm}
  \caption{The workflow of the parent and child process with (a) default fork, (b) shared page table (SPT)-based fork, (c) the proposed Async-fork in the snapshot procedure.}
  \label{fig:relatedwork}
\end{figure*}

A common data persistent approach is to take a point-in-time snapshot of the in-memory data with the system call {\it fork} and dump the snapshot into the file system. Figure~\ref{fig:relatedwork}(a) gives an example of the fork-based snapshot method. In the beginning, the storage engine (the parent process) invokes \emph{fork} to create a child process. As \emph{fork} creates a new process by duplicating the parent process, the child process will hold the same data as the parent. Thus, we can ask the child process to write the data into a file in the background but keep the parent process continuous processing queries. Although the storage engine delegates the heavy IO task to the child process, the fork-based snapshot method can incur latency spikes~\cite{trivago, li2018consistent}. Specifically, queries arriving during the period of taking a snapshot (from the start of \emph{fork} to the end of persisting data) can have a long latency because the storage engine runs into the kernel mode and is out-of-service for queries. For example, the query arrives in time $T_0$ in Figure \ref{fig:relatedwork}(a). For brevity, queries arriving during the period of taking a snapshot are called \emph{snapshot queries}, while the others are called \emph{normal queries}. In general, IMKVSes take a snapshot periodically (e.g., Redis by default takes a snapshot every 60 seconds if at least 10000 records are modified~\cite{redis-default-rdb, zhao2021demand}). Consequently, latency spikes for snapshot queries are not rare.

To solve the problem, researchers proposed a variety of snapshot algorithms such as Copy-on-Update~\cite{liedes2006siren, cao2013fault}, Zigzag~\cite{cao2011fast} and Ping-Pong~\cite{cao2011fast}. These algorithms focus on lowering the cost of taking a snapshot by optimizing the occasion of copying data and reducing the amount of data copied. However, a recent study~\cite{li2018consistent} finds that the performance of the fork-based method is generally competitive with these advanced methods, and even outperforms them for write-intensive workloads. Additionally, the implementation of the fork-based method is very simple and requires a small engineering effort. As such, industrial IMKVSes do not adopt these snapshot algorithms from academia and keep the fork-based as the built-in data persistent approach. As a result, IMKVSes still have latency spikes caused by the fork operation, which can harm the service quality of online applications.

In contrast to existing research focusing on optimizing snapshot algorithms, we propose a straightforward approach to address the problem. Specifically, we optimize the fork operation to reduce the long latency incurred by the fork-based snapshot from the operating system (OS) level, while keeping the data persistent mechanism in IMKVSes unchanged.

We first conduct an in-depth study to reveal the impact of the fork operation on query latency. Our profiling results on the default fork operation show that the overhead of \emph{fork} results in a long tail latency (up to hundreds of milliseconds) for snapshot queries, and copying the page table dominates the execution time of \emph{fork}. This motivates us to further investigate the impact of advanced techniques~\cite{hugepage,THP, mccracken2003sharing, dong2016shared, zhao2021demand} that can accelerate the fork operation. We consider two optimization strategies, the huge page~\cite{hugepage,THP} and the shared page table~\cite{mccracken2003sharing, dong2016shared, zhao2021demand}. As the huge page can degrade the performance of IMKVSes~\cite{redisTroubles, keyDBTrobules}, our profiling only involves the shared page table-based {\it fork} (SPT-based {\it fork})~\cite{zhao2021demand}, which is the latest method. The SPT-based {\it fork} proposed to share the page table between the parent process and the child process in a copy-on-write (CoW) manner to reduce the cost of the fork operation. However, the CoW can frequently interrupt the parent process as shown in Figure~\ref{fig:relatedwork}(b). Our profiling results show that frequent interruption can incur non-negligible overhead for snapshot queries although the SPT-based {\it fork} significantly reduces the tail latency compared with the default fork operation. Moreover, our analysis finds that the shared page table introduces the data leakage vulnerability, which potentially leads to an inconsistent snapshot. Thus, both the huge table and shared page table techniques cannot be applied in this scenario.

Motivated by these findings, we propose Async-fork to mitigate the latency spikes for snapshot queries by optimizing the fork operation. Figure \ref{fig:relatedwork}(c) demonstrates the general idea. As copying the page table dominates the cost of the fork operation, Async-fork offloads this workload from the parent process to the child process to reduce the duration that the parent process runs into the kernel mode. This design also ensures that both the parent and child processes have an exclusive page table to avoid the data leakage vulnerability caused by the shared page table.

However, it is far from trivial to achieve in design since the asynchronization operations of the two processes on the page table can result in an inconsistent snapshot, i.e., the parent process may modify the page table, while the copy operation of the child process is in process. To address the problem, we design the proactive synchronization technique. This technique enables the parent process to detect all modifications (including that triggered by either users or OS) to the page table. If the parent process detects that some page table entries will be modified and these entries are not copied, then it will proactively copy them to the child process. Otherwise, these entries must have been copied to the child process and the parent process will directly modify them. In this way, the proactive synchronization technique keeps the snapshot consistent and reduces the number of interruptions to the parent process compared with SPT-based {\it fork}. Additionally, we parallelize the copy operation of the child process to further accelerate Async-fork.

We implement Async-fork in the Linux kernel (both x86 and ARM64). 
Async-fork is integrated into the OS and transparent to IMKVSes.
The technique is also deployed in the online Redis databases in public Clouds\footnote{\url{https://www.alibabacloud.com/product/apsaradb-for-redis}. Last accessed on 2022/11/13.}. Despite that, we conduct experiments on our local machine for the purpose of test flexibility. In the experiments, we select two popular IMKVSes, Redis and KeyDB, and use the Redis benchmark~\cite{redis-bench} and Memtier benchmark~\cite{memtier}. The database instance size is varied from 1GB, 2GB, and 4GB … to 64GB. Although the SPT-based {\it fork}~\cite{zhao2021demand} may lead to data leakage, our experiment involves this method for comparison purposes because of its efficiency. For Redis, our experiment results show that 1) compared with the default {\it fork}, Async-fork reduces the 99\%-ile latency of snapshot queries by 17.57\% (from 0.074ms to 0.061ms) on 1GB instance, 81.76\% (from 0.435ms to 0.079ms) on 8GB instance and 99.84\% (from 991.9ms to 1.5ms) on 64GB instance; and 2) compared with the latest SPT-based {\it fork}, Async-fork reduces the 99\%-ile latency of snapshot queries by 2.87\% on 1GB instance, 39.73\% on 8GB instance and 61.97\% on 64GB instance. We obtain similar results on KeyDB. These results demonstrate the efficacy of the technique proposed in this paper, especially for the large instances that can lead to long latency.

In summary, we make the following contributions in this paper.

\begin{itemize}
    \item We conduct an in-depth study of the impact of the fork operations on the latency of snapshot queries in IMKVSes.
    \item We propose Async-fork that can mitigate the long latency of snapshot queries from the OS level, which is orthogonal to existing research on the problem.
    \item The technique is implemented in the Linux kernel (both x86 and ARM64) and deployed in the online Redis database in public clouds.
    \item We conduct extensive experiments with Redis and KeyDB to evaluate the efficacy of the proposed techniques. 
\end{itemize}
\section{Background} \label{sec:background}

In this section, we first introduce the preliminaries and then discuss the related work.

\subsection{Preliminary}\label{sec:os_basics}

We first briefly review two operating system concepts, \emph{virtual memory} and \emph{fork} that are closely related to this work. As our technique is implemented and deployed in Linux, we introduce these concepts in the context of Linux. Then, we discuss the use cases of \emph{fork} in databases.

\begin{figure}[t!]
	\centering
	\includegraphics[width=0.9\columnwidth]{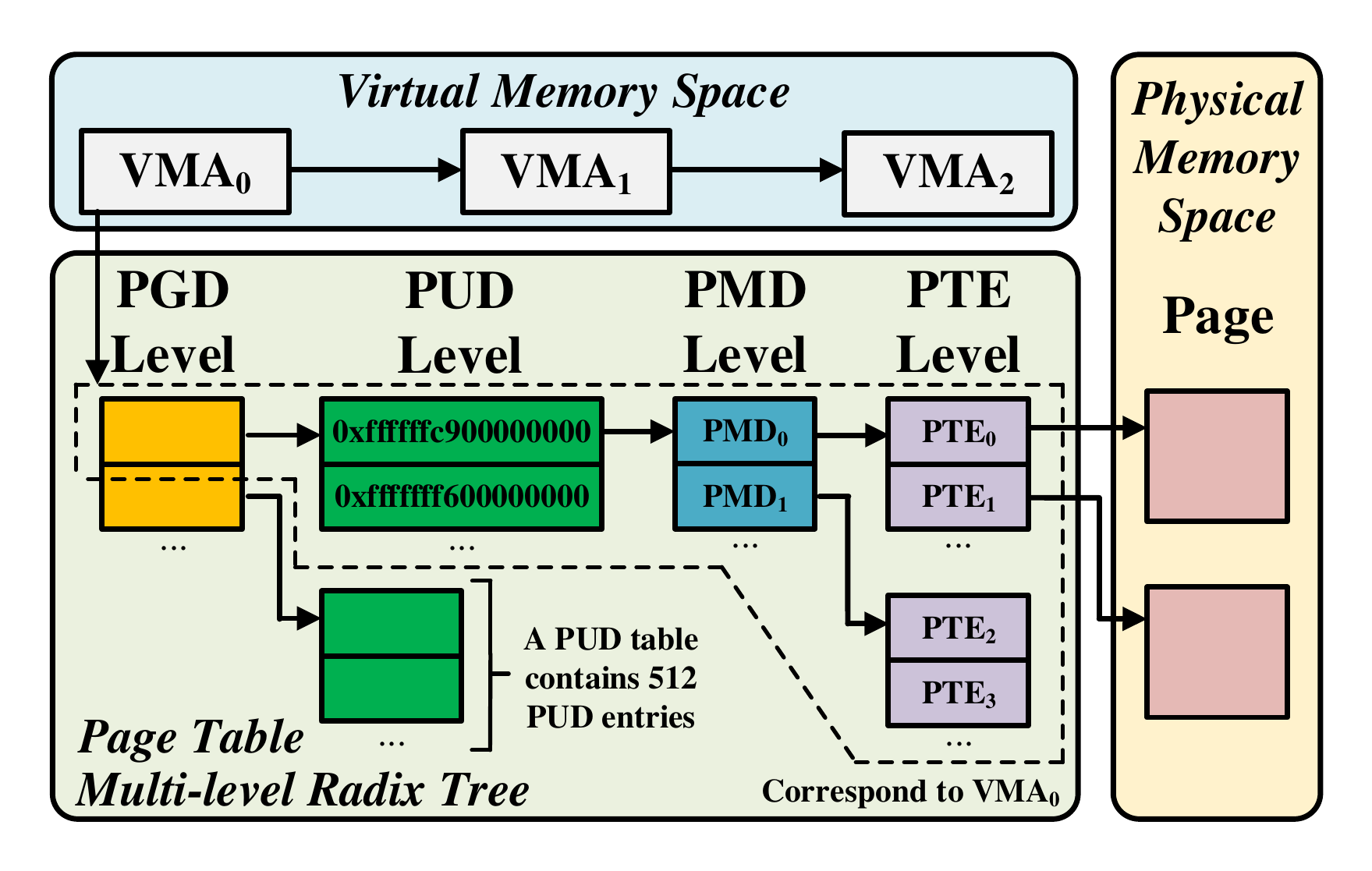}
%	\vspace{-2mm}
	\caption{\sun{The organization of the page table.}}
	\label{fig:page_table}
%	\vspace{-4mm}
\end{figure}

\textbf{Virtual Memory.} As an effective approach to managing hardware memory resources, virtual memory is widely used in modern operating systems. A process has its own virtual memory space, which is organized into a set of virtual memory areas (VMA). Each VMA describes a continuous area in the virtual memory space. The page table is the data structure used to map the virtual memory space to the physical memory. It consists of a collection of page table entries (PTE), each of which maintains the virtual-to-physical address translation information and access permissions. A VMA corresponds to multiple PTEs.

\sun{Figure \ref{fig:page_table} shows an example of the page table.
To reduce the memory cost, the page table is stored as a multi-level radix tree in which PTEs locate in leaf nodes (i.e., PTE Level in Figure~\ref{fig:page_table}). The part in the area marked with the dashed line corresponds to VMA$_0$. The tree at most has five levels. From top to bottom, they are the page global directory (PGD) level, the P4D level, the page upper directory (PUD) level, the page middle directory (PMD) level and the PTE level. As P4D is generally disabled, we focus on the other four levels in this paper. Except for the PTE level, an entry stores the physical address of a page while this page is used as the next-level node (table). With the page size setting to 4KB, a table in each level contains 512 entries. Given a VMA, “VMA’s PTEs” refers to the PTEs corresponding to the VMA and “VMA’s PMDs” is the set of PMD entries that are parents of these PTEs in the tree. For example,
PMD$_0$ and PMD$_1$ belong to VMA$_0$'s PMDs in Figure \ref{fig:page_table}.}

\textbf{Fork Operation.} \emph{Fork} is a system call that creates a new process by duplicating the calling process~\cite{fork2}. Both processes have separate memory spaces. The new (resp. calling) process is called the child (resp. parent) process. To accelerate the operation, Linux implements \emph{fork} with the copy-on-write (CoW) strategy. Specifically, while invoking \emph{fork}, the parent process runs in the kernel mode and copies the metadata (e.g, VMAs, the page table, file descriptors, and signals) to the child process. The PTEs of both parent and child are set to write-protected. After that, the process that first modifies a write-protected page triggers a page fault, which leads to the copy of the page. In a word, benefiting from CoW, \emph{fork} copies the metadata only.

\sun{\textbf{Use Cases of \emph{fork} in Database}. \emph{Fork} has a number of database use cases because it can easily and efficiently create a snapshot of in-memory data, the consistency of which is guaranteed by the OS. In general, these cases can be categorized into two classes based on the usage of the snapshot. First, use \emph{fork} to
delegate dedicated tasks, which have expensive IO or computation costs,
to a child process without blocking the service of the parent process.
MDC~\cite{park2020memory} uses \emph{fork} to record checkpoints for 
in-memory databases. Redis uses \emph{fork} to conduct \emph{log rewriting}~\cite{redis-persist} that optimizes the Append Only File (AOF). FlurryDB~\cite{mior2011flurrydb} proposes to create replica based on \emph{fork} in distributed environments. Second,
use \emph{fork} to create snapshots to support concurrent transaction processing because
\emph{fork} provides the snapshot isolation between processes. HyPer~\cite{kemper2011hyper} proposes to evaluate hybrid OLTP and OLAP queries based on snapshots created by \emph{fork}. AnKer~\cite{sharma2018} designs a fine-grained snapshot mechanism to support MVCC. In particular, AnKer takes a partial snapshot of in-memory data by co-designing the database engine and the system call {\it fork}. Different from our research on accelerating the fork operation from the OS level, AnKer focuses on optimizing which in-memory data should be captured by the snapshot.}

\sun{All these use cases can potentially benefit from Async-fork because 1) they can encounter the query latency spike problem incurred by the fork operation; and 2) Async-fork can accelerate the snapshot creation. This paper focuses on the
scenario that uses \emph{fork} to take a point-in-time snapshot of in-memory data to persist the data~\cite{redis-persist, keydb-persist}. In particular, the storage engine calls \emph{fork} to create a child process that holds the same data as it. Then, the child process writes the data to the hard drive, while the storage engine can continue to serve users’ queries. Although the storage engine delegates the data dump task to the child process, it will be out-of-service for queries during the invocation of \emph{fork} because it runs into the kernel mode. We are particularly interested in
this scenario because the IMKVS is one of the most important services in public cloud and popular IMKVSes (e.g., Redis and KeyDB)~\cite{db-engine}
use this mechanism to persist data. Consequently, these stores encounter serious a query
latency spike problem (see Section \ref{sec:moti}), while they
are generally used in mission-critical applications that have a rigid latency constraint.} \pp{We also evaluate the effectiveness of Async-fork on log rewriting in Redis~\cite{redis-persist}. The experiment results are presented in the technical report~\cite{techreport}.}

\sun{\emph{Remarks.} Instead of developing a general-purpose solution to replace the default {\it fork} in the OS, the goal of Async-fork is to provide an efficient fork operation for the scenarios (especially for IMKVSes) where 1) the applications are memory-intensive, and 2) the parent process is latency-sensitive. Our design allows Async-fork and the default {\it fork} to run in parallel in the OS. Users can easily choose the fork method used in applications (see Section \ref{sec:implementation_details}).}

\subsection{Related Work}\label{sec:related_work}
Consistent snapshot is essential for in-memory databases to support backup and disaster recovery~\cite{zhang2015memory, li2018consistent, diaconu2013hekaton, redis-persistence-demystified}. Some consistent snapshot mechanisms have been proposed to trade off throughput, latency, and memory footprint~\cite{bronevetsky2006recent, cao2013fault, cao2011fast, liedes2006siren, li2018consistent}. Naive snapshot~\cite{bronevetsky2006recent} blocked the storage engine until a deep copy of all the in-memory data is created, which is not suitable for IMKVSes in which the latency is critical. There are also some non-blocking snapshot mechanisms. Copy-on-Update~\cite{liedes2006siren, cao2013fault} proposed to create a shadow copy of the in-memory data; the storage engine is free to access any data but create a deep copy when updates it for the first time. 
Note that, the fork-based snapshot is a Copy-on-Update variant that leverages the operating system. Some other mechanisms used multi-version concurrency control (MVCC)~\cite{bernstein1987concurrency} to keep multiple versions of in-memory data. Zigzag~\cite{cao2011fast} maintained another untouched copy of the in-memory data and introduced metadata bits to indicate which copy the store engine should read from or write to. Based on Zigzag, Ping-Pong~\cite{cao2011fast} maintained three versions of the data to lower the cost of managing metadata bits. Hourglass and Piggyback~\cite{li2018consistent} were developed by combining Zigzag and Ping-Pong.

Although fork-based snapshot results in long latency during snapshot process, popular industrial IMKVSes (Redis and KeyDB) still adopts fork-based snapshot for two reasons: 1) {\it fork} provides a simple engineering implementation for consistent snapshot, while it requires great efforts to integrate the above approaches into the IMKVS. 2) None of the above approaches completely outperform the fork-based snapshot in write-intensive workloads~\cite{li2018consistent}. For example, Ping-Pong and Hourglass can mitigate the latency spikes, while Ping-Pong incurs 3x memory footprint and Hourglass results in higher latency during normal operation. This work resolves the latency spikes of the fork-based snapshot, while keeping its original superiorities.

Previous work~\cite{liedes2006siren, park2020memory} noted that the memory footprint increases during snapshot process due to the CoW strategy. MDC~\cite{park2020memory} proposed to release the pages that have been persisted as soon as possible. AnKer~\cite{sharma2018} introduced a fine-grained version of {\it fork} to take partial snapshot when databases do not need to persist all the data. \sun{CCoW~\cite{electronics11030461} optimized the CoW mechanism based on the spatial locality of memory access. It prioritizes the copy for high-locality memory regions to improve the performance on write-intensive workloads.} Async-fork is orthogonal and complementary to them.

There is also other approach to persist data in IMKVSes. Redis and KeyDB use Append Only File (AOF)~\cite{redis-persist, keydb-persist} to log every write operation received by the storage engine, that will be played again after the database reboots to reconstruct the original dataset. The snapshot and AOF are complementary, and it is recommended to enable both of them simultaneously in IMKVSes.

\section{Motivation}\label{sec:moti}

In this section, we present our profiling results to demonstrate the impact of the fork operation on the query latency. We first evaluate the performance of {\it fork} for taking the snapshot to pinpoint the key performance factors. We then reveal the impact of the fork operation on query latency. In addition to the default {\it fork} in Linux, our profiling involves the state-of-the-art approach~\cite{zhao2021demand} of optimizing the fork operation. Lastly, we summarize our findings according to the profiling results.

\textbf{Profiling Setting.} We use Redis benchmark~\cite{redis-bench} to study the performance in the experiments. The detailed configuration of the test machine is introduced in Section~\ref{sec:experimental-set}.
The experiment reports the latency of a query, that is the elapsed time between the time point that the client issues the query and that the client receives the response.
In the experiment, we enhance the benchmark by generating queries in the open-loop mode~\cite{schroeder2006open, zhang2016treadmill}. This enhancement sends commands to the server without waiting for replies to previous queries to simulate real-world environments. The database instance size is varied from 1GB to 64GB. 
By default, Redis takes one snapshot per 60 seconds if at least 10000 keys changed. In order to measure the impact of the fork operation accurately, we execute the BGSAVE command to trigger the operation of taking a snapshot. We classify queries into two groups, \emph{normal} and \emph{snapshot}, based on their arrival time. 
The snapshot queries are the queries arriving during the period of taking the snapshot (i.e., from the invocation of {\it fork} until the end of persisting the in-memory data), while the others are normal queries. We measure the 99\%-ile (p99) latency and the maximum latency of normal and snapshot queries, respectively. The two latencies greatly impact user experiences and are often used to measure the performance of user-facing databases~\cite{kasture2016, li2018consistent}. We repeat each experiment five times and report the average value.

\subsection{Performance of \emph{fork} for Taking a Snapshot}\label{sec:fork_time}

\begin{figure}
	\centering
	\includegraphics[width=.9\columnwidth]{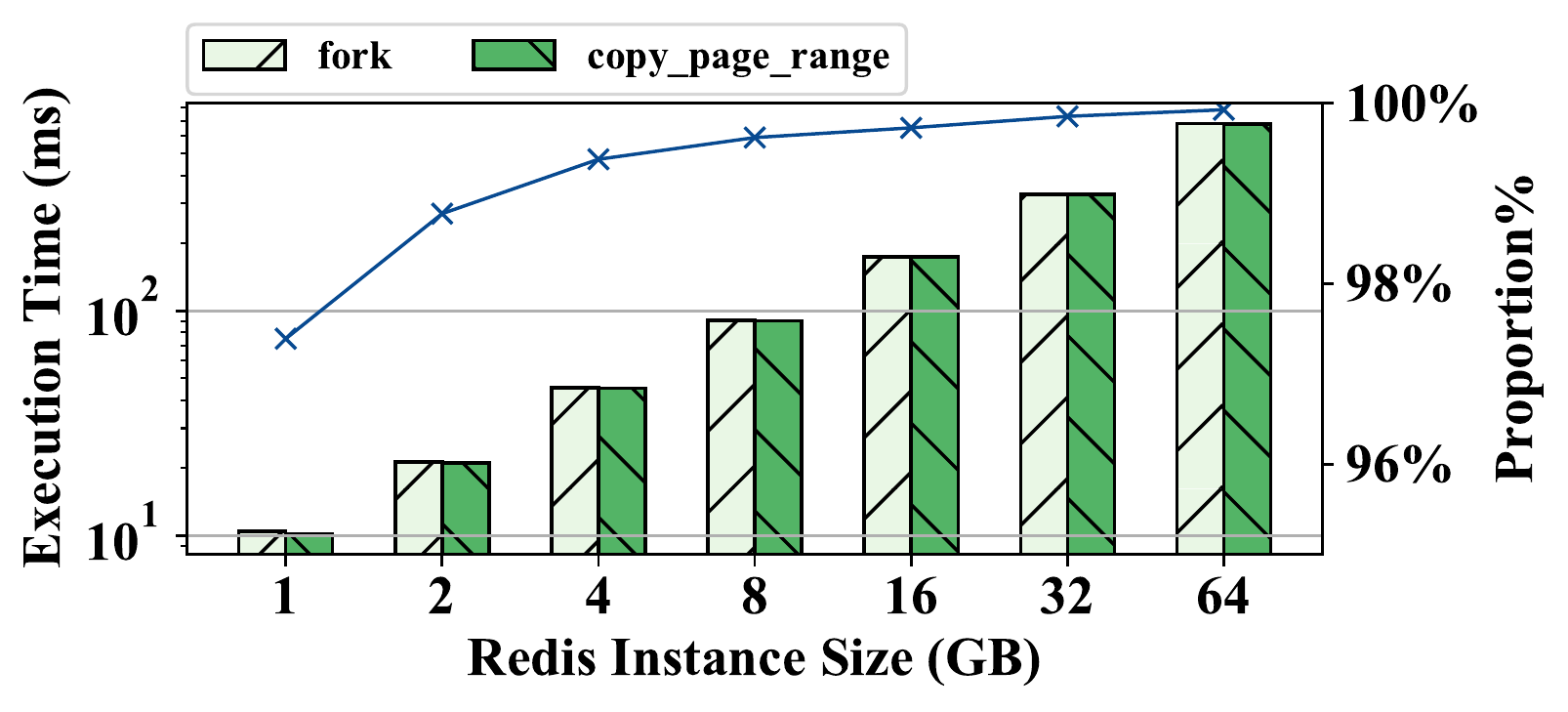}
\caption{The execution time of {\it fork} and the percentage of the time of copying the page table in {\it fork.}}
	\label{fig:characterization_fork}
%		\vspace{-2mm}
\end{figure}

Figure~\ref{fig:characterization_fork} presents the execution time of \emph{fork} and the time of copying the page table in the fork operation. We can see that the execution time grows roughly linearly with the instance size increasing from 1GB to 64GB. The 1GB instance takes less than 10 ms, while the 64GB instance takes more than 600 ms. We can also see that the time of the page table copy dominates the execution time. Particularly, the copy operation takes over 97\% percentage of the execution time on all test cases and up to 99.93\% percentage on the 64 GB instance. Without loss of generality, we measure the detailed metrics of copying the page table on the 8 GB instance to further study the copy operation.

In the experiment, the page table of the 8GB instance has one PGD entry, eight PUDs, $2 ^ {12}$ PMDs and $2 ^ {21}$ PTEs. Overall, \emph{fork} copies the table level-by-level from top to bottom along the radix tree. The copy of one PGD/PUD/PMD entry requires to apply for a page to store its children and initialize the page. The operation takes around 500 ns. Thus, the copy of the $ 2 ^ {12}$ PMDs takes around 2 ms, while the overhead of copying PGDs and PUDs is trivial because there are only a few entries. The rest time (around 70 ms) is spent on copying PTEs. Based on the results, we have the following observation.

\emph{\textbf{Observation 1.} For the fork operation, the execution time dramatically grows with the instance size increasing, and the page table copy dominates the cost. For the page table copy, the overhead of copying PGDs and PUDs is trivial, while that of copying PMDs and PTEs is non-negligible.}

\subsection{Impact of Fork Operations on Latency}\label{sec:problem}

We reveal the impact of the fork operation on the query latency in this subsection. Before presenting the results, we first introduce existing optimization approaches to the fork operation.

\emph{Huge Page}~\cite{hugepage,THP}. In the operating system (OS), we can increase the page size to reduce the number of pages used by a process, for example, setting the page size to 2MB instead of 4KB. The large page size can reduce the number of PTEs and accelerate the fork operation. \sun{However, previous study found that the page fault latency can increase from 3.6$\mu s$ to 378$\mu s$ after enabling huge page because compacting and zeroing memory in the page fault incurs higher overhead on huge page than that on regular page~\cite{kwon2016coordinated}. Moreover, Redis consumes much more memory space after enabling the huge page because applications do not always fully utilize the big page (e.g., in the experiment of~\cite{kwon2016coordinated}, the memory consumption of Redis increased from 12.2GB to 20.7GB). Additionally, allocating huge page can lead to many fragments in the physical memory because it requires consecutive physical memory areas to build a huge page. Consequently, the kernel needs to perform a heavy defragment operation which leads to high CPU utilization~\cite{panwar2018making}. The experiment of~\cite{panwar2018making} shows that the benchmark {\it milc} in SPEC CPU 2006 spends 343 seconds (37\% of its overall execution time) in kernel mode to perform the defragment operation when the memory is highly fragmented.}

\begin{table}[t!]
\caption{Migrating a page from ``X'' to ``Y'' when the parent and child process share the page table.\label{tab:TLB}}
\begin{tabular}{|c|l|l|l|}
\hline
\textbf{Step} & \multicolumn{1}{c|}{\textbf{Operation}}                              & \multicolumn{1}{c|}{\textbf{Parent(P)}}                           & \multicolumn{1}{c|}{\textbf{Child(C)}}                            \\ \hline
1             & Initial state                                                        & \begin{tabular}[c]{@{}l@{}}TLB: V $\rightarrow$ X\\ PTE: V $\rightarrow$ X\end{tabular} & \begin{tabular}[c]{@{}l@{}}TLB: V $\rightarrow$ X\\ PTE: V $\rightarrow$ X\end{tabular} \\ \hline
2             & \begin{tabular}[c]{@{}l@{}}P: Set PTE $\rightarrow$\\ None present\end{tabular} & \begin{tabular}[c]{@{}l@{}}TLB: V $\rightarrow$ X\\ PTE: V $\rightarrow$ N\end{tabular} & \begin{tabular}[c]{@{}l@{}}TLB: V $\rightarrow$ X\\ PTE: V $\rightarrow$ N\end{tabular} \\ \hline
3             & P: Flush TLB                                                         & \begin{tabular}[c]{@{}l@{}}TLB: N/A\\ PTE: V $\rightarrow$ N\end{tabular}    & \begin{tabular}[c]{@{}l@{}}TLB: V $\rightarrow$ X\\ PTE: V $\rightarrow$ N\end{tabular} \\ \hline
4             & \begin{tabular}[c]{@{}l@{}}C: Skipped \\ because N!=X\end{tabular}   & \begin{tabular}[c]{@{}l@{}}TLB: N/A\\ PTE: V $\rightarrow$ N\end{tabular}    & \begin{tabular}[c]{@{}l@{}}TLB: V $\rightarrow$ X\\ PTE: V $\rightarrow$ N\end{tabular} \\ \hline
5             & P: Update PTE                                                        & \begin{tabular}[c]{@{}l@{}}TLB: N/A\\ PTE: V $\rightarrow$ Y\end{tabular}    & \begin{tabular}[c]{@{}l@{}}\textcolor{black}{TLB: V $\rightarrow$ X}\\ \textcolor{black}{PTE: V $\rightarrow$ Y}\end{tabular} \\ \hline
6             & P\&C: Access V                                                       & \begin{tabular}[c]{@{}l@{}}TLB: V $\rightarrow$ Y\\ PTE: V $\rightarrow$ Y\end{tabular} & \begin{tabular}[c]{@{}l@{}}\textcolor{black}{TLB: V $\rightarrow$ X}\\ \textcolor{black}{PTE: V $\rightarrow$ Y}\end{tabular} \\ \hline
\end{tabular}
\end{table}

\sun{Consequently, huge page is recommended to be disabled in many databases (e.g., Couchbase, MongoDB, KeyDB and Redis~\cite{couchbase, mongodb, keyDBTrobules, redisTroubles}). In particular, two No-SQL databases, Couchbase and MongoDB, recommend users to disable the technique because huge page performs poorly with random memory accesses in workloads~\cite{couchbase, mongodb}. KeyDB and Redis, which are two IMKVSes, recommend users to disable huge page because the technique can incur a big latency penalty and big memory usage~\cite{keyDBTrobules, redisTroubles}. Specifically, if we enable the huge page, the parent and child processes share huge pages after calling \emph{fork} to persist on disk. In a busy instance, a few event loops in either of the two processes will cause to target thousands of pages and trigger the copy operation of a large amount of process memory because of the copy-on-write mechanism in OS. Consequently, this leads to a big latency and a big memory usage.} As such, our profiling does not involve this technique.

\emph{Shared Page Table}~\cite{mccracken2003sharing, dong2016shared, zhao2021demand}. This technique proposed to share the page table between parent and child in a copy-on-write (CoW) manner. Specifically, the fork operation returns immediately after copying the metadata except for the page table. The page table will be copied in a CoW manner. %that is introduced in Section \ref{sec:os_basics}.
\sun{However, we find that the shared page table design introduces the data leakage problem, the working set size estimation problem and the NUMA problem. First, the inconsistency between the shared page table and the translation lookaside buffer (TLB) of the child process can lead to the data leakage problem. Second, we cannot accurately estimate the working set size, which indicates the memory usage of each process and is important for cloud resource management~\cite{zhang2020ursa}. This is because the usage is calculated based on the states in the page table~\cite{wss}, while the table is shared by multiple processes. Third, the shared page table and the corresponding processes can locate on different NUMA nodes, which increases the TLB miss overhead~\cite{achermann2020mitosis, panwar2021fast}. Moreover, the NUMA balance mechanism cannot work as expected due to the shared page table. Due to space limit, we discuss the working set size problem and the NUMA problem in the technical report~\cite{techreport}. In the following, we use the example in Table \ref{tab:TLB} to demonstrate the data leakage problem, which is the most serious one among the three problems. 
We also write a test program\footnote{\pp{\url{https://doi.org/10.5281/zenodo.7189585}, Last accessed on 2022/11/13.}} to reproduce the example in practice.}

TLB is the hardware to accelerate the translation from a virtual address to a physical address by caching recent translation results.
Initially, the virtual address “V” is mapped to the physical address “X” in PTE and the mapping is cached in TLB. Note that PTE is shared between parent and child, whereas the two processes have their own TLB entries. Suppose that the memory management mechanism (e.g., memory compaction~\cite{kcompactd, kcompactd-code}, swap~\cite{swap} and NUMA balance~\cite{copypagerange}) of OS mitigates the page from “X” to a new page frame “Y” in the parent process. Then, OS sets the mapping from “V->X” to “V->N” (None Present) to indicate that the mapping is invalid in Step 2 and flushes the parent’s TLB entry in Step 3. 
\sun{For other processes, the OS loops over each of them to check whether its PTEs contain
"V->X"; if so then set the value to "V->N" and flush the TLB; otherwise, skip the process. This works well if each process has a private page table. However, ODF uses the shared page table design. When the OS checks the child process, the PTE has been set to "V->N" in the parent process because the PTE is shared between the parent and child processes and therefore the OS cannot find any PTE with the value "V->X" in the child process. Thus, the OS skips the update to the child in Step 4.} In Step 5, the parent updates PTE to map “V” to “Y”. Although PTE has the correct mapping, the child’s TLB entry is inconsistent with PTE. Consequently, the future access to “V” in the child can lead to a data leakage problem. Despite that, we study the performance of the shared page table-based {\it fork} in our experiments for comparison purposes. \sun{As ODF \cite{zhao2021demand} is the latest work and the only one that is publicly available\footnote{\pp{\url{https://github.com/rssys/on-demand-fork}, Last accessed on 2022/11/13.}} among the methods \cite{mccracken2003sharing, dong2016shared, zhao2021demand} adopting the shared page table design, our experiments focus on ODF.}

\begin{figure}[t]
	\centering
	\includegraphics[width=.9\columnwidth]{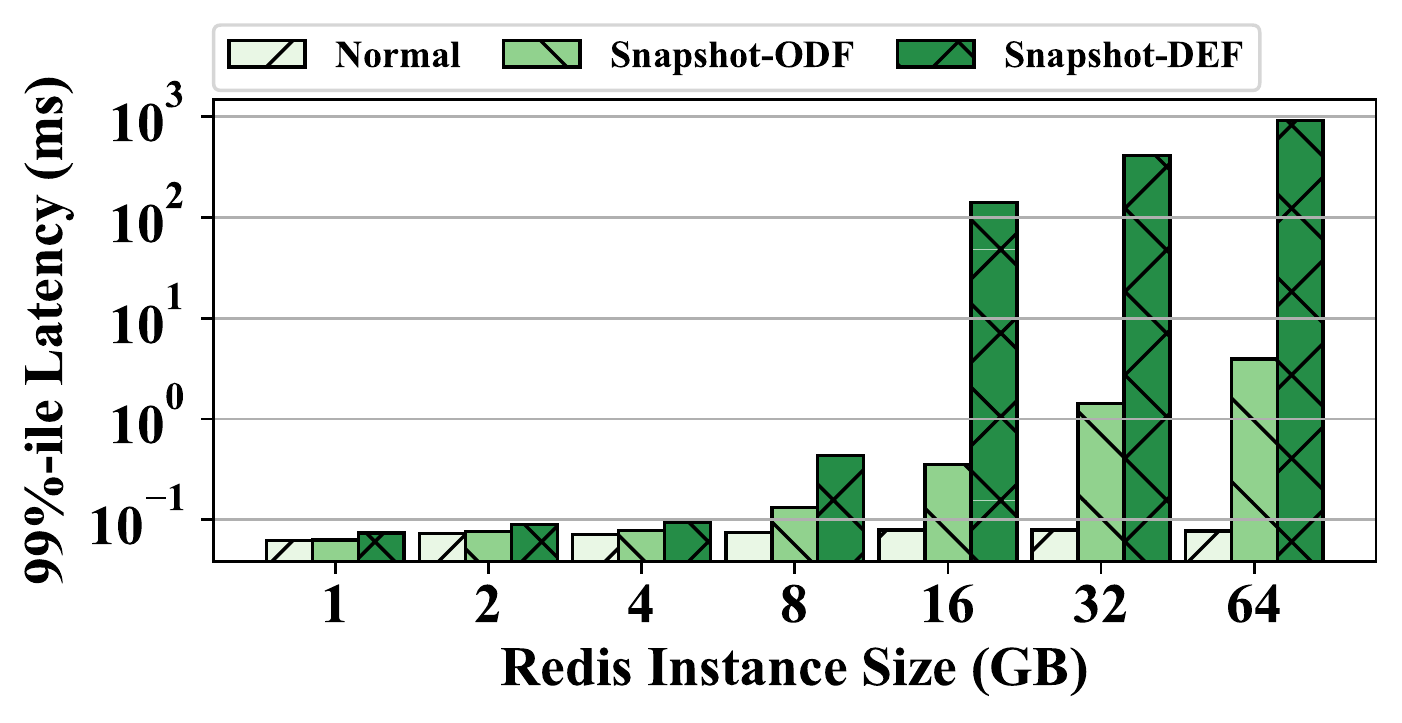}
	\caption{The 99\%-ile latencies of normal queries and snapshot queries in Redis.}
	\label{fig:characterization_p99}
\end{figure}

\begin{figure}[t]
	\centering
	\includegraphics[width=.9\columnwidth]{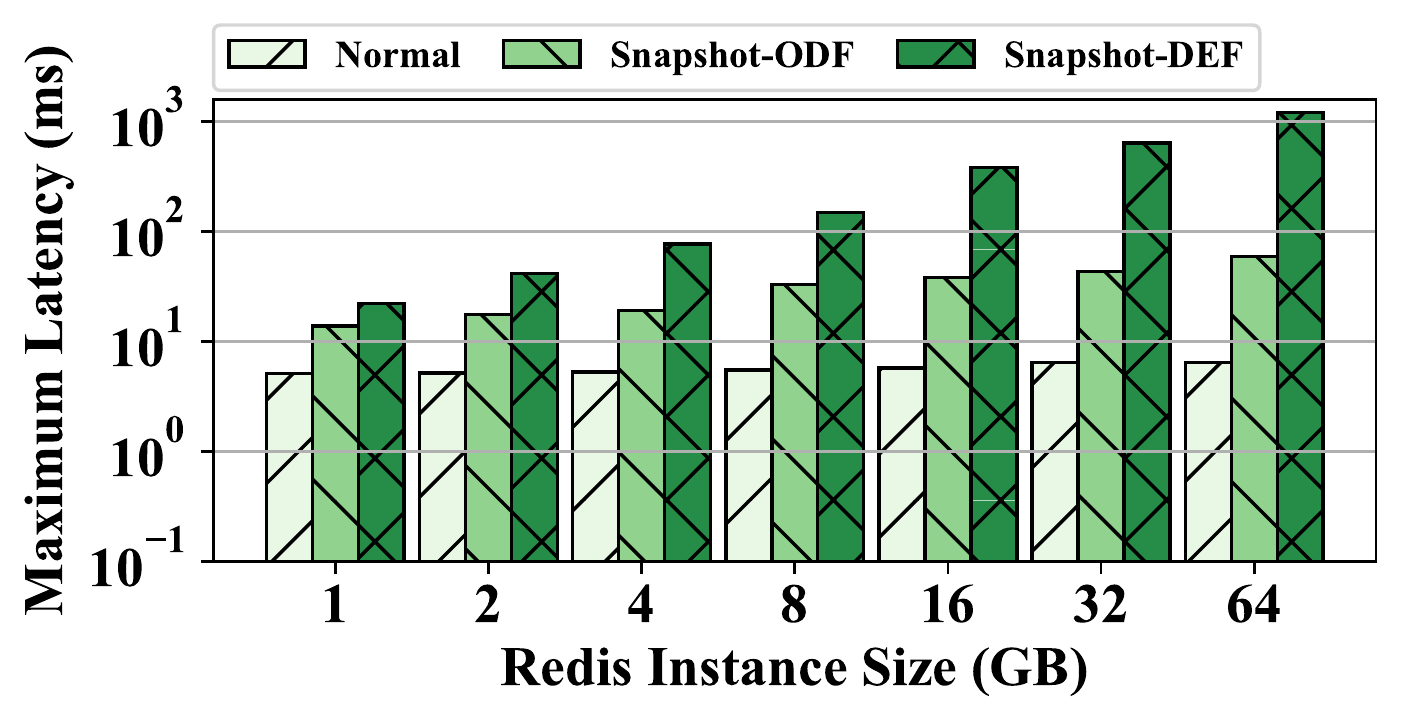}
%	\vspace{-3mm}
	\caption{The maximum latencies of normal queries and snapshot queries in Redis.}
	\label{fig:characterization_max}
%	\vspace{-2mm}
\end{figure}

\textbf{Experiment Results.} Figures \ref{fig:characterization_p99} and \ref{fig:characterization_max} present the 99\%-ile latency and the maximum latency of normal queries and snapshot queries, respectively. Snapshot-ODF denotes the results of using the On-Demand-Fork~\cite{zhao2021demand} to take the snapshot, which is the latest shared page table-based fork method. Snapshot-DEF denotes the results of using the default {\it fork} in Linux. As shown in the figures, the latency of normal queries slightly increases with the instance size varying from 1GB to 64GB. In contrast, the value of snapshot queries grows sharply. The shared page table technique dramatically reduces the latency of the default {\it fork}, especially for the large instance. For example, on the 64GB instance size, the optimization reduces the 99\%-ile latency from 911.95ms to 3.96ms and the maximum latency from 1204.78ms to 59.28ms. The latency of Snapshot-ODF is higher than that of Normal because the CoW of the page table frequently interrupts the engine process. For example, the engine process is interrupted over 7000 times on the 16GB instance, which leads to frequent out-of-service for queries. According to the analysis of existing optimization methods and the experiment results, we have the following observation.

\emph{\textbf{Observation 2.} The fork operation has a significant impact on the latency of snapshot queries, and the tail latency of the default function is up to hundreds of milliseconds. Although the shared page table technique reduces the latency of the default fork operation, the overhead incurred by frequent interruptions is non-negligible and the shared page table introduces the potential data leakage problem.}

\subsection{Summary}

Based on the observations, we find that copying the page table dominates the execution time of the default {\it fork} in Linux, especially, for large instances. The overhead of \emph{fork} results in a long latency (up to hundreds of milliseconds) for queries issued during the invocation of \emph{fork}. Although several optimization methods~\cite{hugepage,THP,mccracken2003sharing, dong2016shared, zhao2021demand} of the fork operation have been proposed, they either lead to poor performance of IMKVSes, or have a data leakage problem, which potentially generates an inconsistent snapshot. Thus, these optimizations cannot be used in the fork-based snapshot mechanism in IMKVSes. As IMKVSes have a rigid requirement on the latency to serve online scenarios, a high-performance {\it fork} is required to reduce the long latency of snapshot queries. 
\section{Design of Async-Fork} \label{sec:design}
In this section, we introduce the design of Async-fork, an operating system-based solution that effectively reduces the long latency of snapshot queries without incurring extra vulnerabilities.

\subsection{General Idea}

\begin{figure}[t]
	\centering
	\includegraphics[width=1\columnwidth]{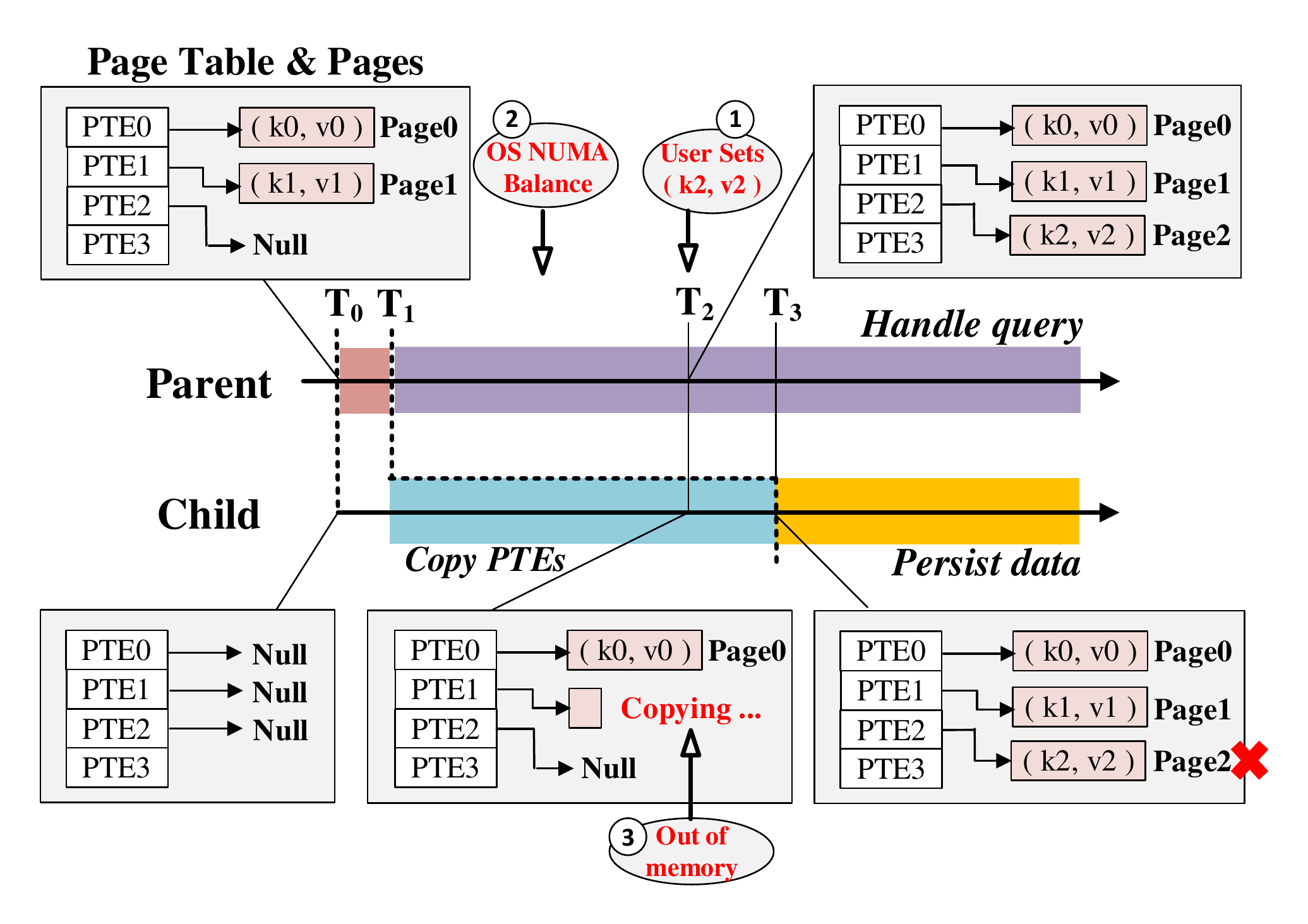}
%	\vspace{-3mm}
	\caption{The challenges in Async-fork.}
	\label{fig:inconsistent}
%	\vspace{-4mm}
\end{figure}

Figure~\ref{fig:relatedwork}(c) shows the general idea of {\bf Async-fork}. 
While the parent process is responsible to copy page tables during snapshot in the default {\it fork}, as shown in the figure, Async-fork offloads the work of copying page table to the idle child process while keeping other steps in the fork unchanged. 
In this way, the parent process is able to handle queries while the child process copies the page table from the parent process simultaneously. 
However, it is non-trivial to achieve the design of Async-fork, as a snapshot may be inconsistent due to the asynchronous operations on the page table.

The inconsistency happens when the parent process modifies a PTE before the child process has copied it.
Take Figure~\ref{fig:inconsistent} as an example. The IMKVS takes a snapshot at time T$_0$, and the in-memory data is $\{(k_0, v_0), (k_1, v_1)\}$. Suppose a user query that sets a new KV pair $(k_2, v_2)$ arrives at time T$_2$ (\ding{172} in Figure~\ref{fig:inconsistent}), and the child process is copying PTEs from time T$_1$ to T$_3$. 
The parent process handles this query, 
and PTE$_2$ is modified to point to a new page (Page$_2$) that contains $(k_2, v_2)$. 
If the child process has not copied the original PTE$_2$ before the modification, 
it would copy the modified PTE$_2$, and owns the new pair $(k_2, v_2)$ in its memory space.
In this way, the data $(k_2, v_2)$ is persisted, and inconsistency happens (the key-value pair $(k_2, v_2)$ does not exist when the snapshot is taken at time $T_0$).

Async-fork resolves the above inconsistency problem by using the parent process to proactively synchronize the modified PTEs to the child process. We explain the detailed steps in Section~\ref{sec:consistency}.
Two main challenges have to be resolved in this solution. 

\begin{algorithm}[t!]
%\SetAlgoNoLine
\footnotesize
\caption{The framework of Async-fork}\label{algo:async-fork}
    \tcp{The parent process}
    \For {each VMA$_i$ in (VMA$_0$, VMA$_1$, VMA$_2$...)} {
        Copy VMA$_i$ to child process\;
        Copy VMA$_i$'s PGDs/PUDs to child process\;
        Set all VMA$_i$'s PMDs to be write-protected\;
    }
    Put the child process on a CPU\ to run it\;
    \While {true} {
        \If {PTE modification is detected and this PTE's PMD is write-protected} {
            Copy PMD to child process\;
            Copy 512 PTEs of this PMD to child process\;
            Set the PMD to be writeable\;
            Set the 512 PTEs to be write-protected\;
        }
    }
    \tcp{The child process}
    \For {each VMA$_i$ in (VMA$_0$, VMA$_1$, VMA$_2$...)} {
        \For {each PMD$_j$ of VMA$_i$} {
            \If {PMD$_j$ is write-protected} {
                Copy PMD$_j$ from parent process\;
                Copy 512 PTEs of PMD$_j$ from parent process\;
                Set the PMD to be writeable\;
                Set the 512 PTEs to be write-protected\;
            }
        }
    }
\end{algorithm}

Firstly, it is necessary to detect all the PTE modifications. However, besides of the user operations, many inherent memory management operations in the operating system also cause PTE modifications. 
For instance, the OS periodically migrates pages among NUMA nodes~\cite{copypagerange}, causing the involved PTEs to be modified as inaccessible (\ding{173} in Figure~\ref{fig:inconsistent}). We describe the method to detect the PTE modifications in Section~\ref{sec:detect}.

Secondly, errors may occur during Async-fork. For instance, the child process may fail to copy an entry due to out of memory (\ding{174} in Figure~\ref{fig:inconsistent}).
In this case, error handling is necessary, as we should restore the process to the state before it calls Async-fork.
We present how errors are handled in Section~\ref{sec:exception}.

By resolving the above challenges, the time of the parent process used on {\it fork} is greatly reduced, and the latency of the snapshot queries can be reduced in consequence. Moreover, after the child process finishes copying the page table, the two processes have their complete private page table. \sun{Therefore, the Async-fork does not introduce the data leakage vulnerability.}

In the following sections, we use the terminology in x86 Linux to explain our design. We also implement  Async-fork on Arm Linux using similar ideas, while similar results are achieved compared with x86. Algorithm~\ref{algo:async-fork} describes how the parent process (Line 1 to 14) and the child process (Line 15 to 24) work in Async-fork.

\subsection{Proactive Synchronization}\label{sec:consistency}

Before introducing the proactive synchronization in detail, we first introduce how we offload the work of copying page table to the child process in Async-fork.

\begin{figure}[t!]
	\centering
	\includegraphics[width=1\columnwidth]{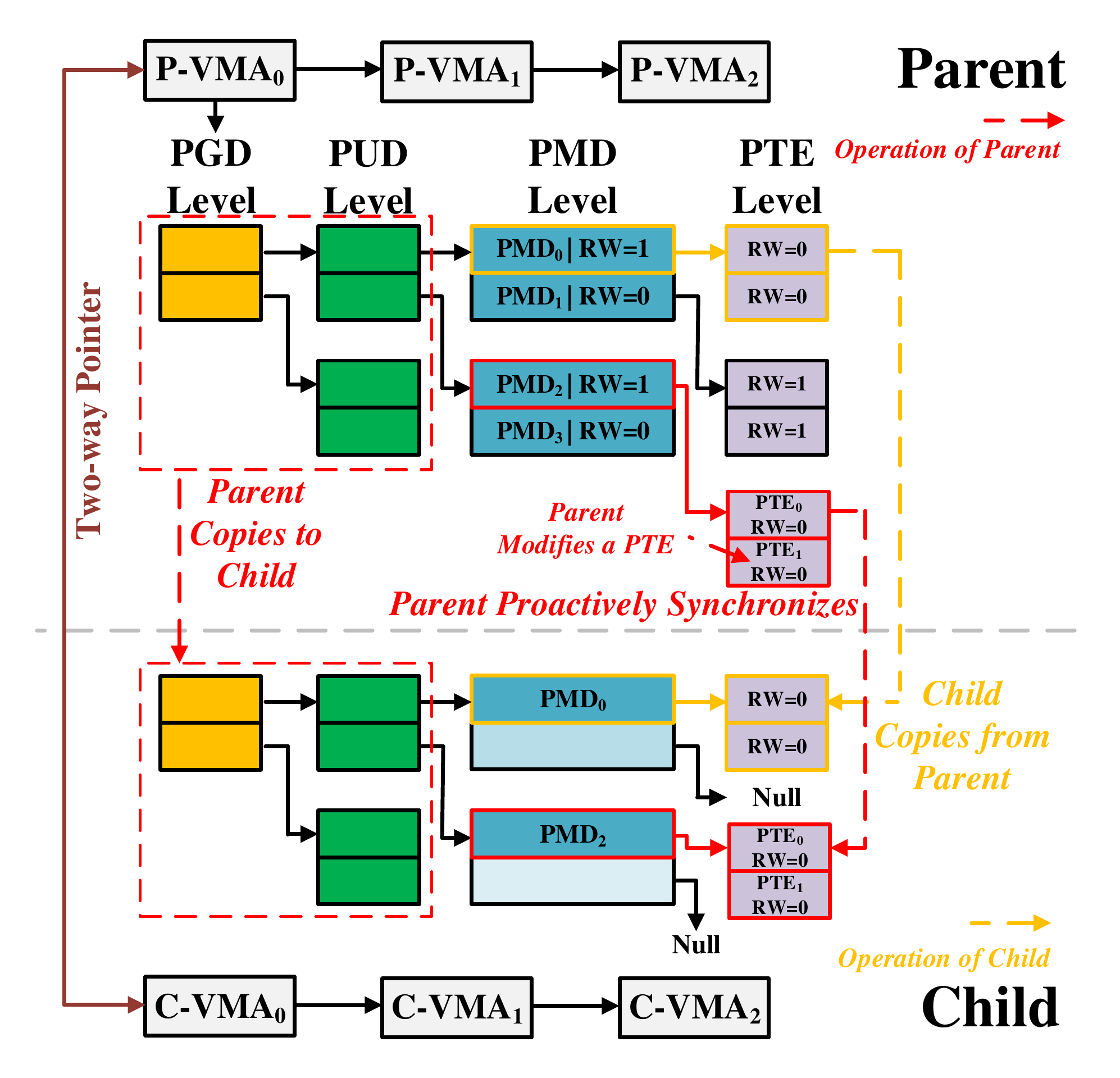}
%	\vspace{-4mm}
	\caption{An example of copying page table in Async-fork. ``RW=1'' represents writable and ``RW=0'' represents write-protected.}
	\label{fig:vmbasics}
%	\vspace{-4mm}
\end{figure}

{\bf Copying Page Table Asynchronously.} In the default {\it fork}, the parent process traverses all its VMAs and copies the corresponding parts of the page table to the child process. The page table is copied from top to bottom. In Async-fork, the parent process roughly follows the above process, but only copies PGD, P4D (if exists) and PUD entries to the child process (Lines 1 to 3 in Algorithm~\ref{algo:async-fork}). After that, the child process starts to run, and the parent process returns to the user mode to handle queries (Lines 7 to 14). 
The child process then traverses the VMAs, and copies PMD entries and PTEs from the parent process. 
Figure~\ref{fig:vmbasics} shows an example of the 
asynchronous page table copy. 
In the figure, the PGD/PUD entries have been copied by the parent process, but PMD entries and PTEs have not yet (some PMD entries point to ``null''). The child process copies PMD entries and PTEs from the parent process (e.g., PMD$_0$ and its PTEs).

We offload the work of copying PMD entries and PTEs to the child process because the overhead of copying them is non-negligible, as analyzed in Section~\ref{sec:fork_time}.
Meanwhile, we keep that the parent process copies PGD/PUD entries. This is because 
the overhead of copying PGD/PUD entries is trivial, 
and it is more robust to minimize the change to the Linux kernel.

{\bf Synchronizing Modified PTEs Proactively.} When the child process is responsible for copying PMD entries and PTEs, it is possible that the PTEs are modified by the parent process before they are actually copied. Note that only the parent process is aware of the modifications (the way to detect the PTE modification will be introduced in Section~\ref{sec:detect}). 

In general, there are two ways to copy the to-be-modified PTEs for the consistency. 1) the parent process proactively copies the PTEs to the child process; 2) the parent process notifies the child process to copy the PTEs and waits until the copying is finished. As both ways result in the same interruption in the parent process, we choose the former way (Lines 8 to 12 in Algorithm~\ref{algo:async-fork}).

More specifically, when a PTE is modified during snapshot, the parent process copies not only this PTE but also all the other PTEs of a same PTE table (512 PTEs in total), as well as the parent PMD entry to the child process proactively. 
For instance, when PTE$_1$ in Figure~\ref{fig:vmbasics} is modified, the parent process proactively copies PMD$_2$, PTE$_0$ and PTE$_1$ to the child process. 
We choose to copy the entire PTE table because we can quickly detect a range of PTEs that will be modified, but accurately identifying which one will be modified is expensive in practice.

{\bf Eliminating Unnecessary Synchronizations.} Always letting the parent process copy the modified PTEs is unnecessary, as it is possible that the to-be-modified PTEs have already been copied by the child process. 
We identify if the PMD entries and PTEs have been copied by the child process, to avoid unnecessary synchronizations. A flag is required to track this status. \sun{We reuse the R/W flag of the PMD entry to record the status. Since the R/W flag is only used when the PMD entry points to a huge page in the x86 Linux kernel, the flag tracks the status correctly. We reuse the R/W flag because 1) this design can avoid adding new fields to the kernel data structures; and 2) the popular databases (e.g., Redis, KeyDB, MongoDB, and Couchbase) recommend disabling the huge page to improve the performance~\cite{redisTroubles, keyDBTrobules, couchbase, mongodb}. Additionally, it is unnecessary to use Async-fork if applications use the huge page because the applications with huge page do not require PTEs but a small number of PGD/PUD/PMD entries (i.e., the page table is small).}

\sun{An alternative approach is to use an unused bit in the \emph{struct page} as the flag. Specifically, each PMD entry points to a PTE table, while OS maintains a data structure (\emph{struct page}) for each PTE table. The \emph{struct page} has bits that are not used by the current Linux kernel. Previous research uses these bits as flags, for example, ODF uses some bits in \emph{struct page} as a reference counter. However, this approach requires further modification to the kernel to initialize the bit. Therefore, we do not adopt the design using the \emph{struct page}.}

If a PMD entry and its 512 PTEs have not been copied to the child process, the PMD entry will be set as write-protected (e.g., PMD$_1$ in Figure~\ref{fig:vmbasics}). Note that, it does not break the CoW strategy of {\it fork} since it still triggers the page fault when the corresponding page is written on x86~\cite{guide2011intel}. Once the PMD/PTEs have been copied to the child process (e.g., PMD$_0$), the PMD entry is changed to be writable (the PTEs are changed to be write-protected to maintain the CoW strategy). Since both parent and child processes lock the page of the PTE table with {\it trylock\_page()} when they are copying PMD entries and PTEs, they will not copy PTEs pointed by the same PMD entry at the same time.

\subsection{Detecting Modified PTEs}\label{sec:detect}

The operations that modify PTEs in the OS can be divided into two categories: 1) {\it VMA-wide modification.} Some operations act on specific VMAs, including creating, merging, deleting VMAs and so on. The modification of a VMA may also cause the VMA's PTEs be modified. For example, the user sends queries to delete lots of KV pairs. The IMKVS (parent process) then reclaims the corresponding virtual memory space by {\it munmap}. Some VMAs are hence split or deleted while the VMAs' PTEs are deleted as well. A VMA is usually large because the operating system always tends to merge adjacent VMAs. It means that the VMA-wide modification usually causes extensive PTE modifications, while many VMA's PMD entries are involved. 
2) {\it PMD-wide modification.} Other operations modify the PTE directly. For example, the page of parent process can be reclaimed by the out of memory (OOM) killer. In this case, one PMD entry is involved. Note that swapping or migrating a 4KB page will change the PTE but the data will not be changed, so we will not handle it. \sun{Due to limited space, we summarize the locations where operations in the OS modify VMAs/PTEs as checkpoints in the technical report~\cite{techreport}.}

We implement the detection by hooking the checkpoints. Once a checkpoint is reached, the parent process checks whether the involved PMD entries and PTEs have been copied (by checking the R/W flag of the PMD entry). For a VMA-wide modification, all the PMD entries of this VMA are checked, while only one PMD entry is checked for a PMD-wide modification. The uncopied PMD entries and PTEs will be copied to the child process before modifying them.

If a VMA is large, the parent process may take a relatively long time to check all PMD/PTE entries by looping over each of them. \sun{We therefore introduce a two-way pointer, which helps the parent process quickly determine whether all entries of a VMA have been copied to the child, to reduce the cost. Each VMA has a two-way pointer, which is initialized by the parent process during the invocation of the Async-fork function. The pointer in the VMA of the parent process (resp. the child process) points to the corresponding VMA of the child process (resp. the parent process). In this way, the two-way pointer maintains a connection between the VMAs of the parent and the child. The connection will be closed after all PMDs/PTEs of the VMA are copied to the child. Specifically, if no VMA-wide modification happens during the copy of PMDs/PTEs of the VMA, then the child closes the connection by setting the pointers in the VMAs of both the parent and child to \emph{null} after the copy operation. Otherwise, the parent will synchronize the modification (i.e., copying the uncopied PMDs/PTEs to the child), and close the connection by setting the pointers to \emph{null} after the copy operation. As both parent and child processes can access the two-way pointers, the pointers are protected by locks to keep the state consistent. When a VMA-wide modification occurs, the parent process can quickly determine whether all PMDs/PTEs of a VMA have been copied to the child by checking the pointer's value, instead of looping over all these PMDs. Besides, the pointer is also used in handling errors (see Section~\ref{sec:exception}).}

\subsection{Handling Errors}\label{sec:exception}

Since copying the page table involves memory allocation, some errors may occur during both the default {\it fork} and Async-fork. 
For instance, a process may fail to initialize a new PTE table due to out of memory. 
Such error may only happen in the parent process in the default {\it fork} and has a standard way to handle the error.
However, the copying of page table is offloaded to the child process in Async-fork, such error may happen in the child process, and a method is required to handle such errors. Specifically, we should restore the parent process to the state before it calls Async-fork, to ensure that the parent process will not crash in the future. 

As Async-fork may modify the R/W flags of the PMD entries of the parent process, we roll back these entries to be writable when errors occur in Async-fork.
Errors may occur 1) when the parent process copies PGD/PUD entries, 2) when the child process copies PMD/PTEs, and 3) during a proactive synchronization. 
In the first case, the parent process rolls back all the write-protected PMD entries. In the second case, the child process rolls back all the remaining uncopied PMD entries. After that, we send a signal ({\it SIGKILL}) to the child process. The child process will be killed when it returns to the user mode and receive the signal.
In the third case, the parent process only rolls back the PMD entries of the VMA containing the PMD entry that is being copied. The purpose is to avoid contending for the PMD entry lock with the child process. An error code is then stored into the two-way pointer of the VMA. \sun{Before (and after) copying PMDs/PTEs of a VMA, the child process will check the pointer to see whether there are errors.} If so, then it stops copying PMD/PTE entries immediately and performs the rollback operations that are already described in the second case.

\section{Optimization and Implementation} \label{sec:opt_imp}
In this section, we present the optimization that further improves the performance of Async-fork, and the way to implement Async-fork in Linux.

\subsection{Accelerating Page Table Copy}\label{sec:acceleration}
The parent process is still interrupted when a proactive PTE synchronization is triggered. 
\sun{A straightforward way to reduce the cost of the proactive PTE synchronization is to let the child process to first copy the PTEs potentially modified before other PTEs. However, this method is not practical because the data accessed by user queries are relatively random and the child process cannot determine which PTEs will be modified when the parent invokes Async-fork.}
We therefore propose to minimize the number of proactive PTE synchronizations by reducing the duration of copying the page table.

As VMAs are independent, the kernel threads can totally perform the copy in parallel and obtain near-linear speedup. Therefore, the child process may launch multiple kernel threads to copy PMD/PTEs in multiple VMAs in parallel, so that the copy completes faster. It can effectively reduce the number of proactive PTE synchronization because the synchronization only happens during the period that the child process is copying PMD/PTEs. The probability of triggering a proactive PTE synchronization gets lower when the period becomes shorter. The experiment in Section~\ref{sec:detail_EXP} shows the efficiency of this optimization.

Multiple kernel threads consume CPU cycles. These threads periodically check whether they should be preempted and give up CPU resources by calling {\it cond\_resched()}, in order to reduce the interference on other normal processes.

\subsection{Implementation of Async-fork}\label{sec:implementation_details}

\begin{figure}
	\centering
	\includegraphics[width=1.0\columnwidth]{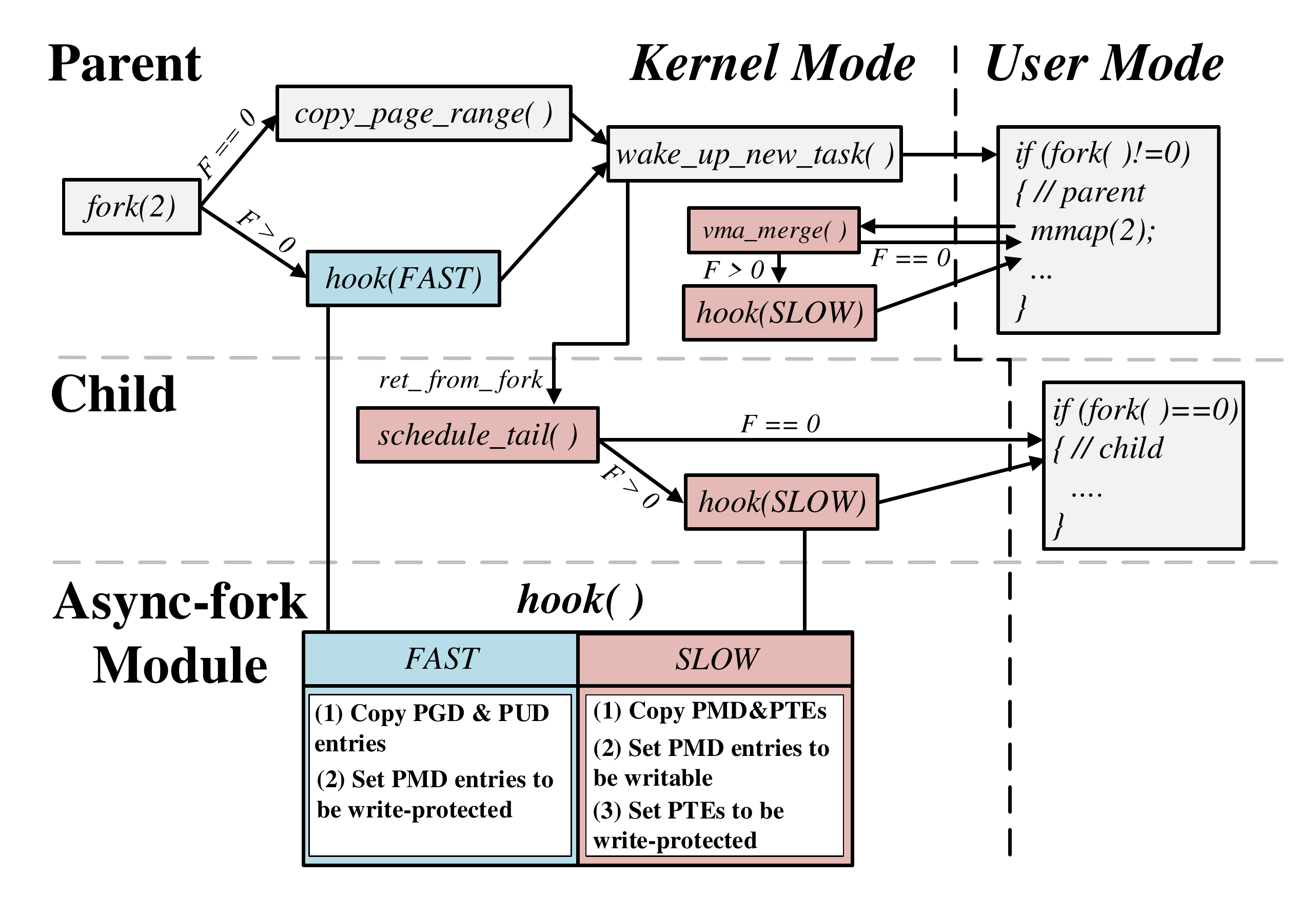}
%	\vspace{-4mm}
	\caption{\sun{The implementation of Async-fork.}}
	\label{fig:acf_implementation}
%	\vspace{-4mm}
\end{figure}

\sun{Figure~\ref{fig:acf_implementation} shows the implementation of Async-fork with a modular design. We encapsulate the code ingested to the kernel into
a hook function, which is instantiated in a kernel module. We can insert/remove the module in/from the kernel as necessary. The hook function is inserted to the call path of default {\it fork} and memory subsystem in the Linux kernel (version 4.19). Users can determine the usage of the default \emph{fork} and Async-fork with the parameter $F$.}

\sun{The hook function is enhanced from {\it copy\_page\_range()}, which accepts a flag (Fast or Slow) to control the copying of the page table. }
With ``Fast'' flag, the function copies PGD/PUD entries and sets PMD entries to be write-protected. 
With ``Slow'' flag, the function copies the write-protected PMD entries and PTEs; when the copying finishes, it sets the PMD entries to be writable and sets the PTEs to be write-protected. When Async-fork is called, the parent process copies the page table to the child process using the function with ``Fast'' flag. 
At the end of Async-fork's invocation, the parent process puts the child process into the {\it runqueue} of a CPU and returns to the user mode. Before the child process returns to the user mode, it copies the page table using the function with ``Slow'' flag. The parent process proactively copies the uncopied PMD/PTEs (using the function with ``Fast'' flag) to the child process before modifying them.

{\bf Flexibility.} %Different applications have different needs in {\it fork}. 
For the IMKVS workload that has small memory footprint, the page table copy is already short. In this case, Async-fork brings small benefit. For these workloads, we provide an interface in {\it memory cgroup} to control whether Async-fork is enabled (as well as the number of kernel threads used to speed up copying PMD/PTEs in the child process). \sun{Specifically, when users add a process to a \emph{memory cgroup}, they can pass a parameter to enable/disable Async-fork at run time (i.e., the parameter {\it F} in Figure~\ref{fig:acf_implementation}). As shown in the figure, if the parameter value is 0, then the process will use the default \emph{fork}. Otherwise, Async-fork is enabled. As such, users can determine which fork operation the process uses as necessary and use Async-fork without any modification in the source code of applications. The process will use the default {\it fork} if no parameter is passed in.}

{\bf Memory overhead.} The only memory overhead of Async-fork comes from the added pointer (8B) in each VMA of a process. In the case, the memory overhead in a process is the number of VMAs times $8B$. Considering a machine with 512GB main memory while 400 processes run simultaneously, there are roughly 760,000 VMAs according to our statistics. In this case, the memory overhead will be 760000$\times$8B $\approx$ 6MB. This overhead is generally negligible.

{\bf Support for ARM64.} The design of Async-fork can also be implemented on ARM64. Specifically, we use the {\it APTable[1:0]}~\cite{arm64} in the table descriptor of the PMD entry as the R/W flag. Async-fork can also be implemented on other architectures that support hierarchical attributes in the page table.

{\bf Consecutive Snapshots.} It is possible that the parent process starts the next snapshot using Async-fork before the previous child process finishes copying PMD/PTEs, as the parent process returns to the user mode before the child process. 
In the current implementation, we will not block the next Async-fork call but keep a VMA's page table be copied by only one child process at any time. When the parent process copies a VMA to the child process, it checks the two-way pointer to identify whether there exists a previous child process copying the page table of this VMA. If exists, it proactively copies the whole page table of this VMA to the previous child process. \sun{Async-fork adopts this design because supporting concurrent fork operations need to track all child processes
that are copying page tables. This is hard because the kernel does not provide the information
and we need to inject new data structures into the kernel to record the states and synchronize these
child processes.}

\sun{As Async-fork cannot support concurrent fork operations in a process (the same
as the default {\it fork} in the OS), Async-fork cannot support the cases where
the parent process needs to conduct the fork operation in an ultra-high
frequency (e.g., in milliseconds). In spite of the limitation, Async-fork works well for database use cases because a storage engine does not create many child processes from the parent process simultaneously in practice. Specifically, conducting the fork operation frequently leads the parent to frequently turn into the kernel mode, which hurts the service quality. Moreover, many processes executing in parallel will degrade the performance due to the resource contention (e.g., IO bandwidth and CPU resources).
Therefore, IMKVSes (e.g., Redis and KeyDB) do not recommend ultra-frequent data snapshots (generally 60 seconds~\cite{redis-default-rdb}) and have no cases requiring the parent process to invoke Async-fork in milliseconds to our knowledge. For HyPer, which uses the fork operation to support concurrent transaction processing, Async-fork can work well. This is because the parent process handles OLTP (a single updater) that has a rigid requirement on the latency, whereas the child process executes OLAP which generally has a long execution time and is more tolerant of the latency than OLTP. Moreover,
HyPer notices the cost of the fork operation and designs a novel mechanism,
which makes multiple OLAP queries to share a snapshot, to improve the performance.}

\section{Evaluation} \label{sec:evalution}
\label{sec:evaluation}

In this section, we evaluate the effectiveness of Async-fork.

\subsection{Experimental Setup} \label{sec:experimental-set}
We evaluate Async-fork on a machine with two Intel Xeon Platinum 8163 processors, each of them has 24 physical cores (48 logical cores). 
The machine has 384GB memory, and 1TB NVMe hard drive. 
In terms of software setup, the experimental platform runs CentOS 7.9 with Linux 4.19.
Except for the experiments in Section~\ref{sec:ECS}, all the other experiments are conducted within a single machine as described above.

\textbf{Benchmarks.} We use Redis (version 5.0.10)~\cite{redis} and KeyDB (version 6.2.0)~\cite{keydb} compiled with gcc 6.5.1 as the representative IMKVS servers, and use Redis benchmark~\cite{redis-bench} as well as Memtier benchmark~\cite{memtier} to be the workload generators. 
The benchmarks reveal the scenario where multiple clients send requests to the IMKVS server simultaneously. Similar to prior work~\cite{chen2019parties}, we enhance Redis benchmark to generate queries in an open-loop mode for measuring the latency accurately~\cite{schroeder2006open, zhang2016treadmill}.

By default, the experiments are conducted with the following settings: 1) 50 clients (default settings) are used in Redis benchmark, while 50 clients are used in Memtier benchmark for consistency. 
2) The key range is set to $2\times 10^{8}$, the key size of 8B and the value size of 1024B. 3) Each experiment is repeated by five times and the average results are reported. 4) In Async-fork, the child process launches 7 additional kernel threads (together with the child process itself, there are 8 threads in total) to help it copy PMDs/PTEs faster. Our experiment in Figure~\ref{fig:no-multi-threaded-copy} shows that Async-fork still outperforms the state-of-the-art solution with a single child process.
5) The KeyDB server is configured with 4 threads.

\textbf{Metrics.} We launch a large number of queries ($5\times 10^{6}$) to the IMKVS and record the latency of the queries that arrive in the snapshot process (start from the parent process calls {\it fork} until the child process persists all in-memory data). \sun{Specifically, we measure the 99\%-ile latency because latency-sensitive services generally provide an SLA on some percentile. Moreover, we report the maximum latency because IMKVS is often used for demanding use cases that have a rigid requirement for the worst-case latency~\cite{redis-maxlatency}. For example, the increase of the maximum latency of Redis can lead to read error on connections~\cite{trivago}. Therefore, in the production environment~\cite{tairpmem}, the maximum latency is an important indicator of system stability. Due to limited space, we report the experiment results of query processing
throughput and total out-of-service time of the parent process in the technical report~\cite{techreport}. In the report, we also discuss the method of tuning IMKVSes to further improve the performance.}

\textbf{Baselines.} We compare Async-fork with On-Demand-Fork (denoted by {\it ODF} in short)~\cite{zhao2021demand}, the state-of-the-art shared page table-based fork. In ODF, each time a shared PTE is modified by a process, not only one PTE but 512 PTEs located on the same PTE table will be copied at the same time. 
We do not report the results of the default {\it fork} in this section since it results in 
10X higher latency compared with both Async-fork and ODF in most cases (already presented in Section~\ref{sec:problem}). To accurately measure the performance of the IMKVSes in the experiments, we trigger the snapshot operation manually using the BGSAVE command.

\subsection{Overall Evaluation}\label{sec:exp_write_intensive}

We first evaluate Async-fork using write-intensive workloads that require frequent snapshots for data persistence. 

\begin{figure}[t!]
	\centering
	\includegraphics[width=0.9\columnwidth]{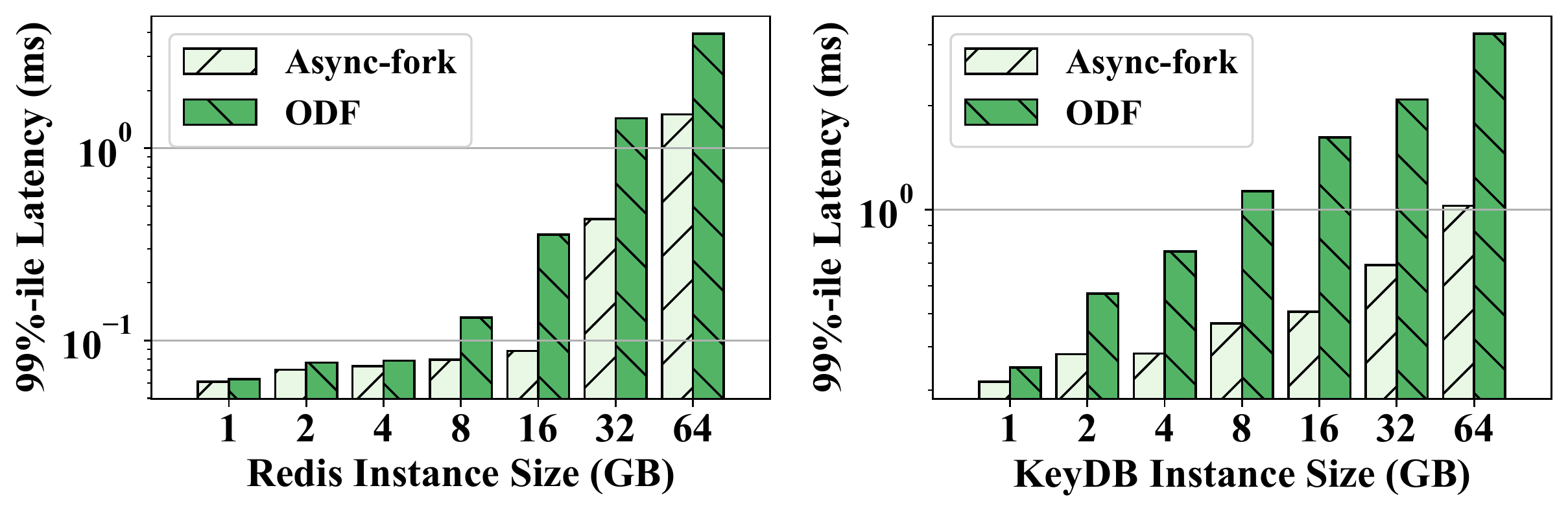}
	\caption{The 99\%-ile latency of snapshot queries.}
	\label{fig:write-p99}
%	\vspace{-2mm}
\end{figure}

\begin{figure}[t!]
	\centering
	\includegraphics[width=0.9\columnwidth]{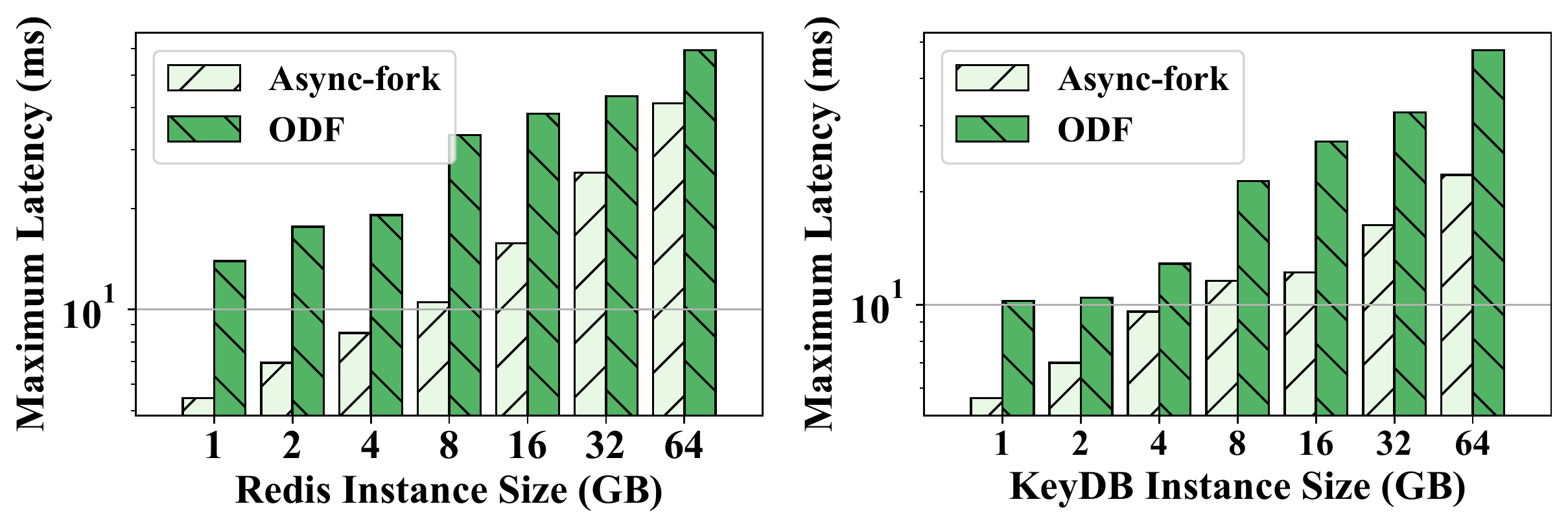}
	\caption{The maximum latency of snapshot queries.}
	\label{fig:write-max}
%	\vspace{-2mm}
\end{figure}

{\bf Latency Results.} In this experiment, we use the Redis benchmark~\cite{redis-bench} to generate the write-intensive workload by issuing {\it SET} queries to the IMKVSes, and configure the clients to send 50,000 such queries in a second.
Since the IMKVS is often used in the context of demanding use cases, there are usually strict requirements on both the 99\%-ile latency and the worst case latency~\cite{redis-maxlatency, keydb-maxlatency}.

Figure~\ref{fig:write-p99} shows the 99\%-ile latencies of snapshot queries in Redis and KeyDB, with ODF and Async-fork.
As observed, Async-fork outperforms ODF in all the cases, and the performance gap increases when the instance size gets larger. 
For instance, operating on a 64GB IMKVS instance, the 99\%-ile latency of the snapshot queries is 3.96ms (Redis) and 3.24ms (KeyDB) with ODF, while the 99\%-ile latency reduces to 1.5ms (Redis, 61.9\% reduction) and 1.03ms (KeyDB, 68.3\% reduction) with Async-fork.

Figure~\ref{fig:write-max} shows the maximum latency of the snapshot queries.
As observed, Async-fork greatly reduces the maximum latency of the benchmarks compared with ODF, even if the instance size is small.
For a 1GB IMKVS instance, the maximum latencies of the snapshot queries are 13.93ms (Redis) and 10.24ms (KeyDB) respectively with ODF, while the maximum latencies are decreased to 5.43ms (Redis, 60.97\% reduction) and 5.64ms (KeyDB, 44.95\% reduction) with Async-fork.

\begin{figure}[t!]
	\centering
	\includegraphics[width=0.9\columnwidth]{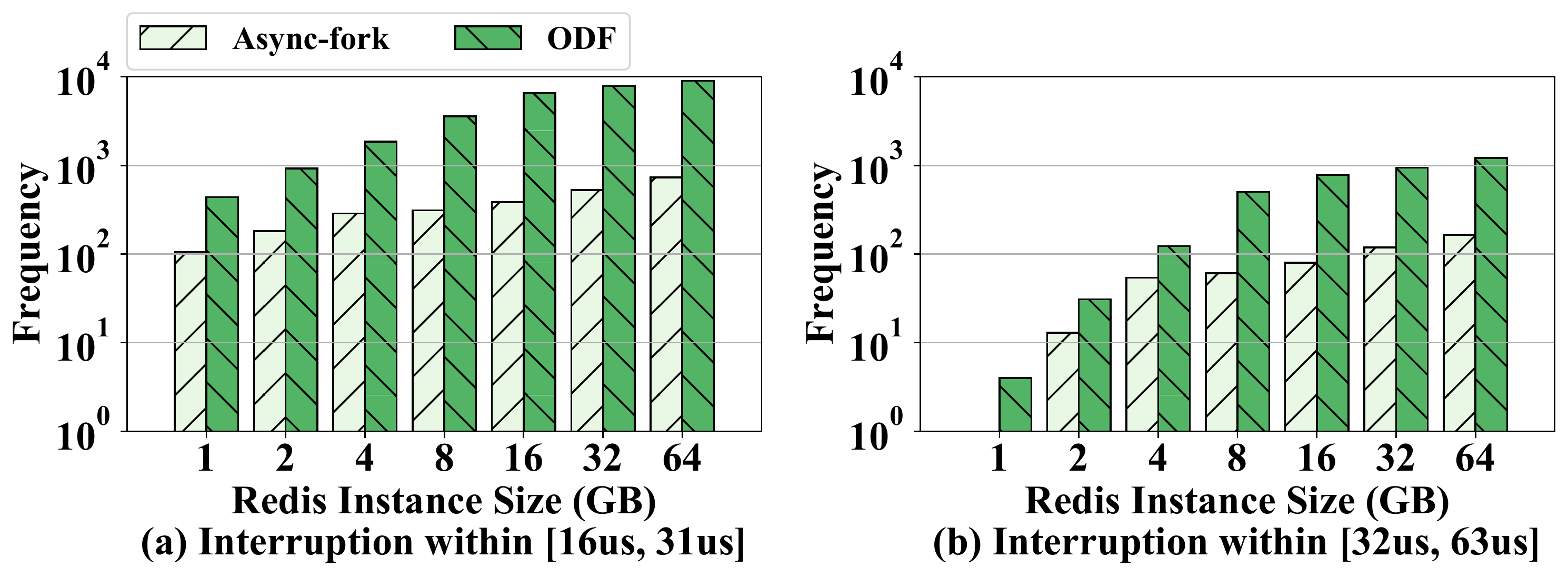}
	\caption{The frequency of interruptions in the parent process during the snapshot process.}
	\label{fig:pmd_copy}
%	\vspace{-2mm}
\end{figure}

{\bf Deep Diving.} \sun{The interruption of Async-fork is caused by the proactive synchronization, whereas the interruption of ODF is caused by the CoW of the page table.
We measure the interruptions of the parent process during the snapshot to understand the reason that Async-fork outperforms ODF. Specifically, the parent process turns into kernel mode and is out of service for queries when executing the \emph{copy\_pmd\_range()} function. The overhead of \emph{copy\_pmd\_range()} dominates the cost of executing in the kernel mode.
In order to examine the number of interruptions and the out-of-service time, we use the \emph{bcc} tool~\cite{bcc} to count the number of \emph{copy\_pmd\_rage()} invocations and measure the execution time of each invocation. The result of \emph{bcc} is a histogram in which the bucket is the time duration and the frequency is the number of invocations whose execution time falls into the bucket. The categories [16$us$, 31$us$] and [32$us$, 63$us$] are two default buckets of \emph{bcc}. In our experiments, all invocations fall into the two buckets.}

\sun{Figure~\ref{fig:pmd_copy} shows the frequency of the interruptions within [16us, 31us] and [32us, 63us]. We can see that Async-fork significantly reduces
the frequency of interruptions. For example, Async-fork reduces the frequency of interruptions from 7348 to 446 on the 16GB instance.} Async-fork greatly reduces the interruptions because 
the interruptions happen only when the child process is copying PMD/PTEs (the required time is within 600ms as in Figure~\ref{fig:multi-threaded-copy}). 
However, the interruption can happen until all data is persisted by the child process in ODF, while the data persistence operation requires tens of seconds (e.g., persisting 8GB in-memory data takes about 40s). 
Under the same workload, the parent process is more vulnerable to interruption when using ODF.

\subsection{Detailed Evaluation}{\label{sec:detail_EXP}}		

\begin{figure}[t!]
	\centering
	\includegraphics[width=0.9\columnwidth]{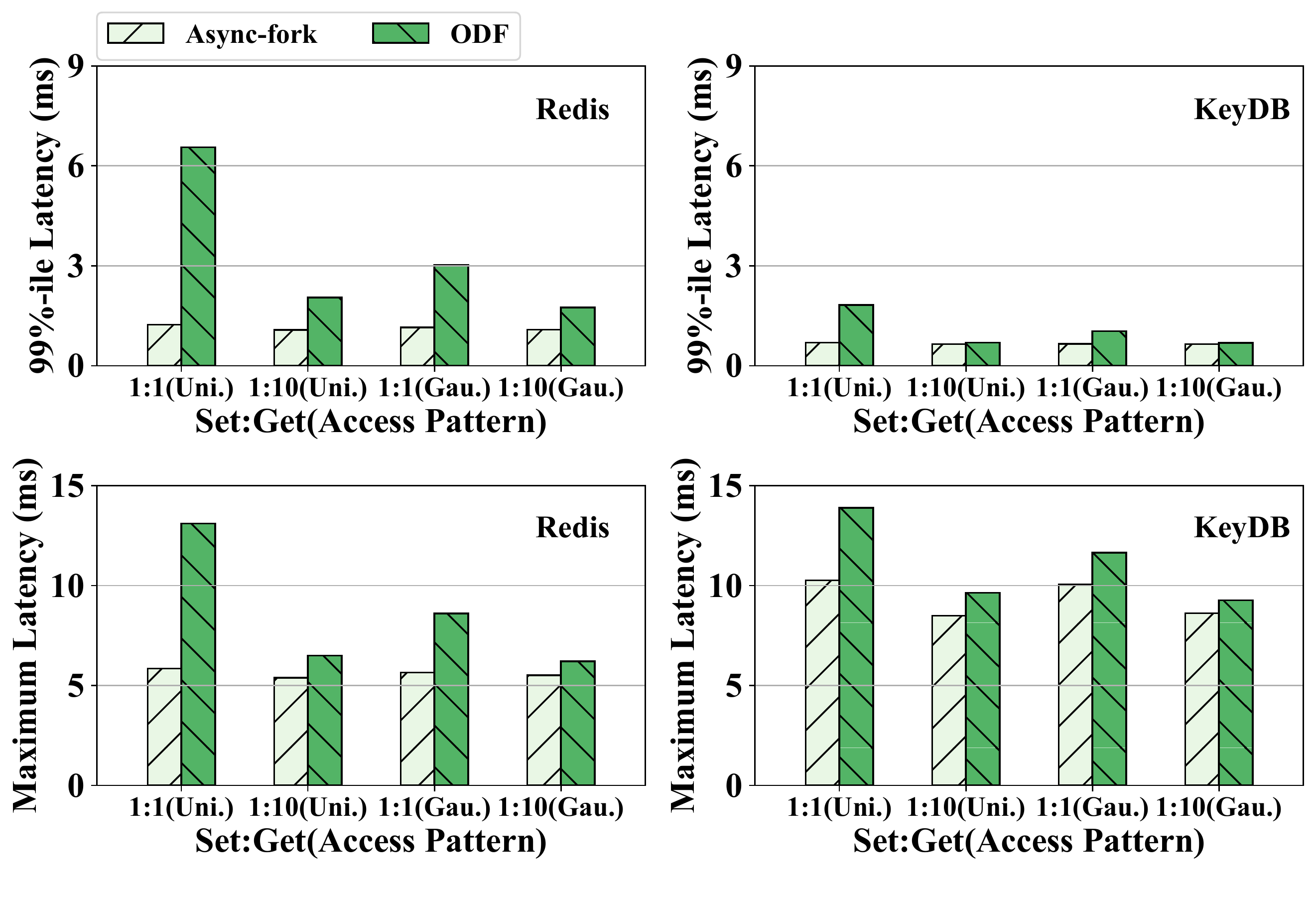}
	\caption{The 99\%-ile latency and maximum latency of snapshot queries under different workloads in an 8GB IMKVS.}
	\label{fig:rw-mix-max}
%		\vspace{-2mm}
\end{figure}

{\bf Sensitivities to the Read/Write Patterns.} %We then evaluate Async-fork using the workloads with different read-write patterns. 
Figure~\ref{fig:rw-mix-max} shows the results using four workloads with different read-write patterns generated with Memtier~\cite{memtier}. 
In the figure,  ``1:1 (Uni.)'' represents the workload with 1:1 Set:Get Ratio and the uniform random access pattern, while ``1:10 (Gau.)'' is the workload with 1:10 Set:Get ratio and the Gaussian distribution access pattern. 

Observed from Figure~\ref{fig:rw-mix-max}, Async-fork still outperforms ODF. 
The benefit is smaller for the workload with more {\it GET} queries. 
This is because the serving (parent) process only copy a small number of PTEs for the {\it GET}-intensive workloads. 
Moreover, the modified memory is smaller in the experiment with the Gaussian distribution compared with the uniform random access pattern. 
With random pattern, the key-value pairs in the IMKVS have the same probability of being accessed, and parts of key-value pairs may be accessed repeatedly with the Gaussian Distribution access pattern. 
Since shared PTEs are copied only when they are modified for the first time in ODF, the parent process is interrupted fewer times 
with the Gaussian distribution access pattern. %The performance of ODF therefore improves.

In general, Async-fork works better for write-intensive workloads. The larger the modified memory is, the better Async-fork performs. \sun{Integrating Async-fork with CCoW~\cite{electronics11030461} to improve the performance on write-intensive workloads is an interesting research direction because Async-fork can utilize CCoW to copy the PTEs of high-locality memory pages in advance to reduce the number of proactive synchronizations.}

\begin{figure}[t!]
	\centering
	\includegraphics[width=0.9\columnwidth]{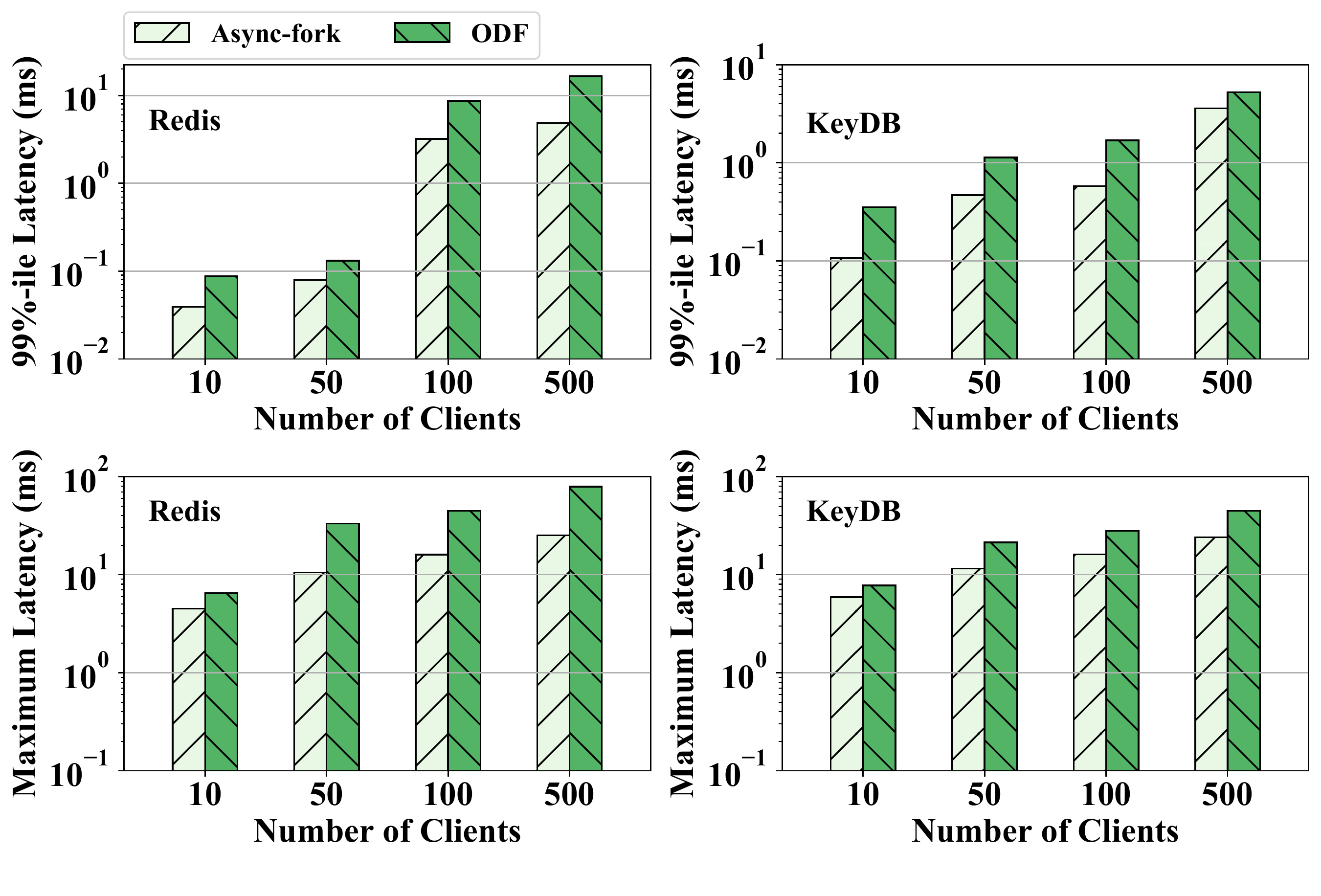}
	\caption{The 99\%-ile latency and the maximum latency of snapshot queries under different numbers of clients in an 8GB IMKVS.}
	\label{fig:write-intensive-client}
%	\vspace{-2mm}
\end{figure}

{\bf The Impact of the Number of Clients.} 
In this experiment, we change the number of clients in Redis-benchmark while keeping sending 50, 000 {\it SET} queries every second to an 8GB IMKVS server. 
Figure~\ref{fig:write-intensive-client} shows the results of 99\%-ile and maximum latency with 10, 50, 100 and 500 clients. As observed, Async-fork outperforms ODF, while the performance gap increases as the number of clients increases. 
This is because more requests arrive at the IMKVS at the same time when the number of clients increases. 
As a result, more PTEs may be modified at the same time, and the duration of one interruption to the parent process may become longer.

\begin{figure}[t!]
	\centering
	\includegraphics[width=0.9\columnwidth]{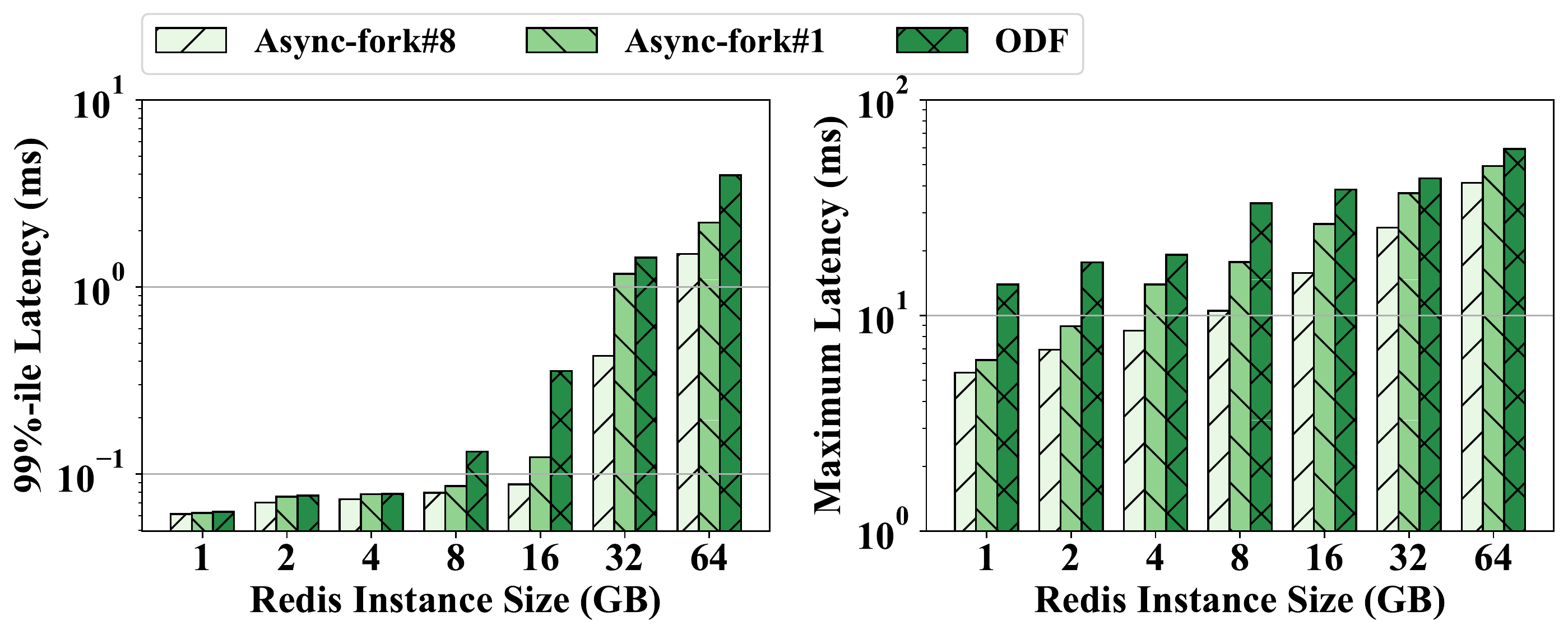}
	\caption{The 99\%-ile and maximum latency of snapshot queries when Async-fork uses 1 or 8 threads.
	}
	\label{fig:no-multi-threaded-copy}
\end{figure}

\begin{figure}[t!]
	\centering
	\includegraphics[width=0.9\columnwidth]{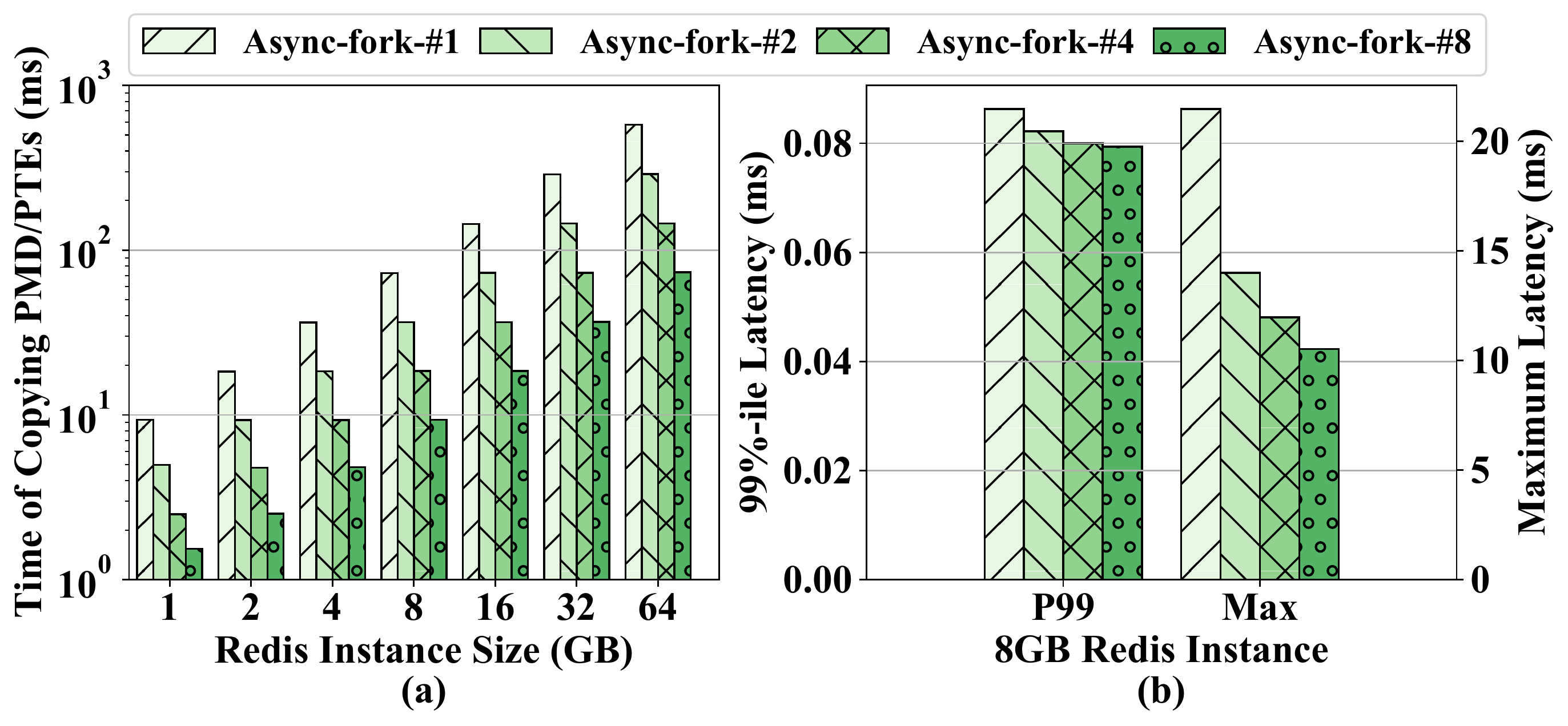}
	\caption{(a) The time that the child process takes to copy PMDs/PTEs in Async-fork. (b) The 99\%-ile and maximum latency in an 8GB Redis instance.}
	\label{fig:multi-threaded-copy}
\end{figure}

{\bf The Impact of the Number of Threads In the Child Process.}  
In Async-fork, multiple kernel threads may be used to copy PMD/PTEs in parallel in the child process.
Figure~\ref{fig:no-multi-threaded-copy} shows the 99\%-ile and maximum latency of snapshot queries under different Redis instance sizes. 
Async-fork\#$i$ represents the results of using $i$ threads in total to copy the PMDs/PTEs.

Observed from Figure~\ref{fig:no-multi-threaded-copy}, 
Async-fork\#1 (the child process itself) still brings shorter latency than ODF. 
The maximum latency of the snapshot queries is decreased by 34.3\% on average, compared with ODF. 
We can also find that using more threads (Async-fork\#8) can further decrease the maximum latency. 
This is because the sooner the child process finishes copying PMDs/PTEs, the lower probability the parent process is interrupted to proactively synchronize PTEs.

In more detail, Figure~\ref{fig:multi-threaded-copy}(a) shows the time that the child process takes to copy PMDs/PTEs with different numbers of kernel threads, while Figure~\ref{fig:multi-threaded-copy}(b) shows the corresponding 99\%-ile and maximum latency in an 8GB Redis instance. 
As we can see, launching more kernel threads effectively reduces the time of copying PMDs/PTEs in the child process. The shorter the time is, the lower the latency becomes.

\begin{figure}[t]
	\centering
	\includegraphics[width=0.9\columnwidth]{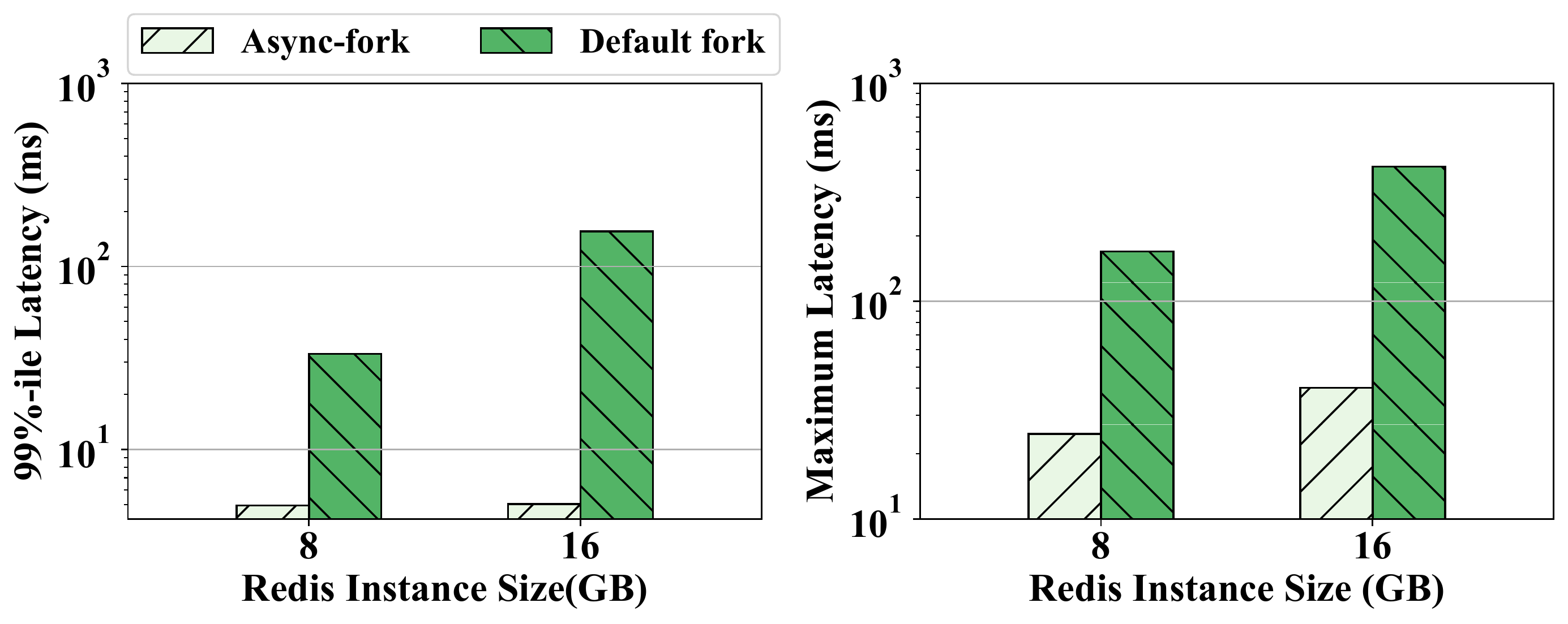}
	\caption{The 99\%-ile latency and maximum latency of snapshot queries with Async-fork in production.}
	\label{fig:ecs}
%	\vspace{-2mm}
\end{figure}

\subsection{Evaluation in Production Environment}\label{sec:ECS}

Async-fork has been deployed in our production Redis Clouds. We rent a Redis database from our public cloud to evaluate Async-fork in the production environment. The rented Redis server has 16GB memory and 80GB SSD. We also rent another virtual machine with 4 vCPU cores and 16GB memory from the same cloud to run the Redis benchmark as the client. The client settings are the same as previous experiments. The network bandwidth between the IMKVS server and the client is 3Gbps. 

We compare Async-fork with the default {\it fork} in this subsection because ODF is not available in our cloud. Figure~\ref{fig:ecs} shows the 99\%-ile latency and maximum latency of the snapshot queries with the default {\it fork} and Async-fork. 
As observed, with an 8GB instance, the 99\%-ile latency of the snapshot queries is reduced from 33.29ms to 4.92ms, and the maximum latency is reduced from 169.57ms to 24.63ms. With a 16GB instance, the former is reduced from 155.69ms to 5.02ms, while the latter is reduced from 415.19ms to 40.04ms.

Before Async-fork is deployed on our Redis Clouds, 
we received a large number of complaints from tenants about the long latency when taking the snapshot. 
Some tenants even have to disable the snapshot function, and they cannot reboot their Redis servers, otherwise the data lost. 
After we deploy Async-fork, no more latency complaints have been received.

\section{Conclusion} \label{sec:conclusion}

In this paper, we study the latency spikes incurred by the fork-based snapshot mechanism in IMKVSes and address the problem from the operating system level. In particular, we conduct an in-depth study to reveal the impact of the fork operation on the latency spikes. According to the study, we propose Async-fork. 
It optimizes the fork operation by offloading the workload of copying the page table from the parent process to the child process. To guarantee data consistency between the parent and the child, we design the proactive synchronization strategy. 
Async-fork is implemented in the Linux kernel and deployed in production environments. Extensive experiment results show that the technique proposed in this paper can significantly reduce the tail latency of queries arriving during the snapshot period.

\begin{acks}
% This work was supported by the [...] Research Fund of [...] (Number [...]). Additional funding was provided by [...] and [...]. We also thank [...] for contributing [...].
This work was partially sponsored by the National Natural Science Foundation of China (62232011, 62022057), and Shanghai international science and technology collaboration project (21510713600). This work was also supported by Alibaba Group through Alibaba Innovative Research Program. Quan Chen and Minyi Guo are the corresponding authors.
\end{acks}

%\clearpage

\bibliographystyle{ACM-Reference-Format}
\bibliography{main}

\appendix
\section{Issues Incurred by Shared Page Table Design}

\sun{We initially consider using ODF in our cloud to solve the query latency spike problem. However, we find that the shared page table design adopted in ODF~\cite{zhao2021demand} (as well as previous studies~\cite{mccracken2003sharing, dong2016shared}) incurs several issues, which block its usage in the production environment.}

\sun{\textbf{Data Leakage Problem.} As presented in Section \ref{sec:problem}, ODF has the data leakage problem due to the inconsistency between the page table and the TLB. A potential fixing method for the data leakage vulnerability works as follows: 1) for each PTE, record the processes that share it; and 2) if the PTE of a page is set to be invalid in the parent process, then the OS notifies the other processes sharing the PTE to flush the TLB entry. However, it is not easy to record all the processes that share a PTE; this problem is similar to the \emph{reverse mapping} problem, i.e., mapping a physical page back to the PTEs that correspond to it. Although the reverse mapping concept looks simple and easy to implement, it is hard in practice due to the space and access efficiency problem. As introduced in the blogs\footnote{\url{https://lwn.net/Articles/23732}. Last accessed on 2022/11/13.}\vphantom{d}$^{\text{,}}$\footnote{\url{https://lwn.net/Articles/383162}. Last accessed on 2022/11/13.} by Jonathan Corbet, the author of Linux Device Drivers, each reverse mapping method requires complicated data structures and an update mechanism. Unfortunately, for each PTE, ODF records the number of processes that share it, but has no reverse mapping information. To that end, ODF cannot support the fixing method.}

\sun{\textbf{WSS Estimation Problem.} The Linux kernel cannot accurately estimate the working set size (WSS) of a process using ODF, while WSS is an important metric, which cloud providers utilize to improve the resource utilization. Specifically, WSS is the memory size that a process actually requires in run time. A process can populate a memory space that is much bigger than the actual usage (e.g., apply for a memory pool with a fixed size at startup but do not shrink in run time). The Linux kernel estimates WSS of a process by counting how many pages the process has visited (based on the accessed bit of the PTE~\cite{wss}). However, the shared page table design in ODF prevents the kernel from distinguishing the visitor of a page. Consequently, the Linux kernel cannot give an accurate estimation on WSS. In practice, it is very important for public clouds to improve resource utilization because tenants generally would like to apply for more resources than their actual demands (e.g., 68.4\% of memory space is wasted in our clouds~\cite{zhang2020ursa}).}

\sun{\textbf{NUMA Problem.} We find that the NUMA balancing mechanism does not work as expected with ODF. Specifically, when the NUMA balancing mechanism decides to change the page location for the child process, it will set PTEs to be PROT\_NONE (i.e., accessing the pages in the future triggers page faults) to ask the child process to handle the faults. However, both the parent and child processes can capture the faults due to the shared page table design. Additionally, the overhead of TLB miss increases when the processes sharing the page table run on different NUMA nodes because there is only one copy of PTEs, and some processes and the PTEs can locate into different nodes. The overhead of TLB miss can increase due to page table walk on the remote node~\cite{achermann2020mitosis, panwar2021fast}. Although our experiments do not consider NUMA environments, we mention this issue to show the gap between ODF and practice.}

\sun{Fixing these issues is far from trivial and requires significant efforts (e.g., adding new data structures and further modifying the kernel). This motivates us to develop a new fork mechanism that can work in the production environment. Different from the shared page table design, Async-fork has the same design as the default fork operation in which both parent and child processes have a private page table. The parent's PTE table is locked if a PTE (belongs to this table) will be updated in the parent process. Therefore, when the PTE is updated in the parent process, either the PTE has been copied by the child process before the update or not. For the first case, both child and parent processes have their own PTEs whereby Async-fork has no data leakage issue. For the second case, the child process will copy the PTE after the update because of the lock on the parent's PTE table. We use the example in Table~\ref{tab:TLB_FCM} to demonstrate this case. The parent's PTE table is locked during the page migration (Step 2 to Step 4). The PTE can be accessed by the child process after Step 4. Then, the child process will copy the updated PTE. As such, Async-fork does not suffer from the data leakage problem.} 

\sun{Nevertheless, the idea that
shares a page table among multiple processes and uses CoW to keep consistent is promising
because it can reduce the workload of copying the page table. It is an interesting
and challenging future research to address all the above issues and develop a fork method based
on the shared page table design.}

\begin{table}[]
\caption{\sun{Migrating a page from ``X'' to ``Y'' before the child process copies the PTE (“V->X”) in Async-fork.\label{tab:TLB_FCM}}}
\begin{tabular}{|c|l|l|l|}
\hline
\textbf{Step} & \multicolumn{1}{c|}{\textbf{Operation}}                                         & \multicolumn{1}{c|}{\textbf{Parent(P)}}                                                        & \multicolumn{1}{c|}{\textbf{Child(C)}}                                                         \\ \hline
1    & Initial state                                                          & \begin{tabular}[c]{@{}l@{}}TLB:V-\textgreater{}X\\ PTE:V-\textgreater{}X\end{tabular} & \begin{tabular}[c]{@{}l@{}}TLB:N/A\\ PTE:N/A\end{tabular}                             \\ \hline
2    & \begin{tabular}[c]{@{}l@{}}P: Set PTE $\rightarrow$\\ None present\end{tabular} & \begin{tabular}[c]{@{}l@{}}TLB:V-\textgreater{}X\\ PTE:V-\textgreater{}N\end{tabular} & \begin{tabular}[c]{@{}l@{}}TLB:N/A\\ PTE:N/A\end{tabular}                             \\ \hline
3    & P: Flush TLB                                                           & \begin{tabular}[c]{@{}l@{}}TLB:N/A\\ PTE:V-\textgreater{}N\end{tabular}               & \begin{tabular}[c]{@{}l@{}}TLB:N/A\\ PTE:N/A\end{tabular}                             \\ \hline
4    & P: Update PTE                                                           & \begin{tabular}[c]{@{}l@{}}TLB:N/A\\ PTE:V-\textgreater{}Y\end{tabular}               & \begin{tabular}[c]{@{}l@{}}TLB:N/A\\ PTE:N/A\end{tabular}                             \\ \hline
5    & C: Copy PTE                                                            & \begin{tabular}[c]{@{}l@{}}TLB:N/A\\ PTE:V-\textgreater{}Y\end{tabular}               & \begin{tabular}[c]{@{}l@{}}TLB:N/A\\ PTE:V-\textgreater{}Y\end{tabular}               \\ \hline
6    & P\&C: Access V                                                            & \begin{tabular}[c]{@{}l@{}}TLB:V-\textgreater{}Y\\ PTE:V-\textgreater{}Y\end{tabular} & \begin{tabular}[c]{@{}l@{}}TLB:V-\textgreater{}Y\\ PTE:V-\textgreater{}Y\end{tabular} \\ \hline
\end{tabular}
\end{table}

\begin{table*}[t!]
\centering
\caption{Operations that may modify VMAs/PTEs.\label{tab:modify}}
%\vspace{2mm}
\footnotesize
\begin{tabular}{|c|c|c|l|c|c|c|}
\hline
\textbf{VMA-wide}                                                                                                                & \textbf{Description}                                                                                                                                                                  & \textbf{Checkpoint}                                                                                                                                                                                        & \multicolumn{1}{c|}{\textbf{}} & \textbf{PMD-wide} & \textbf{Description}                                                                                                                         & \textbf{Checkpoint}               \\ \cline{1-3} \cline{5-7} 
\textit{\begin{tabular}[c]{@{}c@{}}mmap(2),munmap(2),\\ madvise(2),mprotect(2),\\ mlock(2),munlock(2),\\ mremap(2)\end{tabular}} & \begin{tabular}[c]{@{}c@{}}User uses system calls\\ proactively, casuing\\ the properties of VMAs to\\ be modified or VMAs to be\\ deleted, split, extended\\ or merged.\end{tabular} & \textit{\begin{tabular}[c]{@{}c@{}}vma\_merge( ),\_\_split\_vma( ),\\ detach\_vmas\_to\_be\_unmapped( ),\\ madvise\_vma( ),\\ do\_mprotect\_pkey( ),\\ mlock\_fixup( ),\\ vma\_to\_resize( )\end{tabular}} &                                & page fault        & \begin{tabular}[c]{@{}c@{}}A virtual address is\\ accessed for the first\\ time and OS allocates a\\ page that is mapped by it.\end{tabular} & \textit{\_\_handle\_mm\_fault( )} \\ \cline{1-3} \cline{5-7} 
\textit{expand\_stack( )}                                                                                                        & \begin{tabular}[c]{@{}c@{}}User/OS expands the\\ address space of stack.\end{tabular}                                                                                                 & \textit{\begin{tabular}[c]{@{}c@{}}expand\_upwards( ),\\ expand\_downwards( ),\end{tabular}}                                                                                                               &                                & out of memory     & \begin{tabular}[c]{@{}c@{}}OOM killer reclaims\\ pages.\end{tabular}                                                                         & \textit{zap\_pmd\_range( )}       \\ \cline{1-3} \cline{5-7} 
NUMA balance                                                                                                                     & \begin{tabular}[c]{@{}c@{}}OS periodically migrates\\ pages among NUMA nodes.\\ PTEs are modified to\\ be inaccessible.\end{tabular}                                                  & \textit{change\_prot\_numa( )}                                                                                                                                                                             &                                & get user page     & \begin{tabular}[c]{@{}c@{}}Direct I/O, VFIO pin\\ get pages and\\ modifies them.\end{tabular}                                                & \textit{follow\_page\_pte( )}     \\ \hline
\end{tabular}
\end{table*}

\section{Implementation Details}

We summarize the operations that modify VMAs/PTEs in the OS as checkpoints.
Table~\ref{tab:modify} lists the functions (i.e., checkpoints) that can modify a VMA or a PTE, through a comprehensive analysis on the Linux kernel. As shown in the table, some operations are caused by users' queries, while others are caused by OS inherent memory management operations. Different kernel versions show quite stable call paths. \sun{Our analysis shows that only one function related to Async-fork has ever been modified from Linux kernel 2.6 to 6.0 (see Table~\ref{tab:modify-life}). Thus, the kernel related to our work is very stable.}

\sun{In practice, updating the kernel is a critical task for both cloud providers and users. They tend to stay with a kernel version for a relatively long time to keep the stability of services. Therefore, it is common for cloud providers to customize the OS kernel for important applications or specific internal needs in the production environment even the modification to the kernel cannot be upstreamed. For example, Facebook uses a customized disk cache component\footnote{\url{https://engineering.fb.com/2013/10/09/core-data/flashcache-at-facebook-from-2010-to-2013-and-beyond}. Last accessed on 2022/11/13.} in the kernel to speed up MySQL, and Google adds specific rules\footnote{\url{https://lwn.net/Articles/871195}. Last accessed on 2022/11/13.} for the out-of-memory killer in their kernel. The implementation of Async-fork adopts the modular design, which can be easily inserted in (or removed from) the kernel.}

\begin{table}[t!]
\centering
\caption{\sun{The lifecycle and location of related functions.}\label{tab:modify-life}}
\footnotesize
\begin{tabular}{|c|c|c|}
\hline
\textbf{Function}                 & \textbf{Lifecycle} & \textbf{Location} \\ \hline
vma\_merge()                      & v2.6.12 - v6.0     & mm/mmap.c         \\ \hline
\_\_split\_vma( )                 & v2.6.33 - v6.0     & mm/mmap.c         \\ \hline
detach\_vmas\_to\_be\_unmapped( ) & v2.6.12 - v6.0     & mm/mmap.c         \\ \hline
madvise\_vma( )                   & v2.6.12 - v5.16.20 & mm/madvise.c      \\ \hline
do\_mprotect\_pkey( )             & v4.9 - v6.0        & mm/mprotect.c     \\ \hline
mlock\_fixup( )                   & v2.6.12 - v6.0     & mm/mblock.c       \\ \hline
vma\_to\_resize( )                & v2.6.33 - v6.0     & mm/mremap.c       \\ \hline
expand\_upwards( )                & v2.6.15 - v6.0     & mm/mmap.c         \\ \hline
expand\_downwards( )              & v2.6.23 - v6.0     & mm/mmap.c         \\ \hline
change\_prot\_numa( )             & v3.8 - v6.0        & mm/mempolicy.c    \\ \hline
\_\_handle\_mm\_fault( )          & v3.12 - v6.0       & mm/memory.c       \\ \hline
zap\_pmd\_range( )                & v2.6.12 - v6.0     & mm/memory.c       \\ \hline
follow\_page\_pte( )              & v3.16 - v6.0       & mm/gup.c          \\ \hline
\end{tabular}

\end{table}

\section{Supplement Experiments}

In this section, we first report the experiment results of query processing throughput and total out-of-service time for the parent process during the snapshot process. We then report the experiment results of using Async-fork in log rewriting. Additionally, we report the execution time of Async-fork and ODF. Lastly, we discuss the method of tuning IMKVSes to further improve the performance.

{\bf Throughput Results.}
This experiment measures the processing throughput of the IMKVS when taking a snapshot of the in-memory data. To do so, we use Redis benchmark~\cite{redis-bench} to launch a great many queries and report the number of completed queries on the server side as the processing throughput.

\begin{figure}[t!]
	\centering
	\includegraphics[width=0.9\columnwidth]{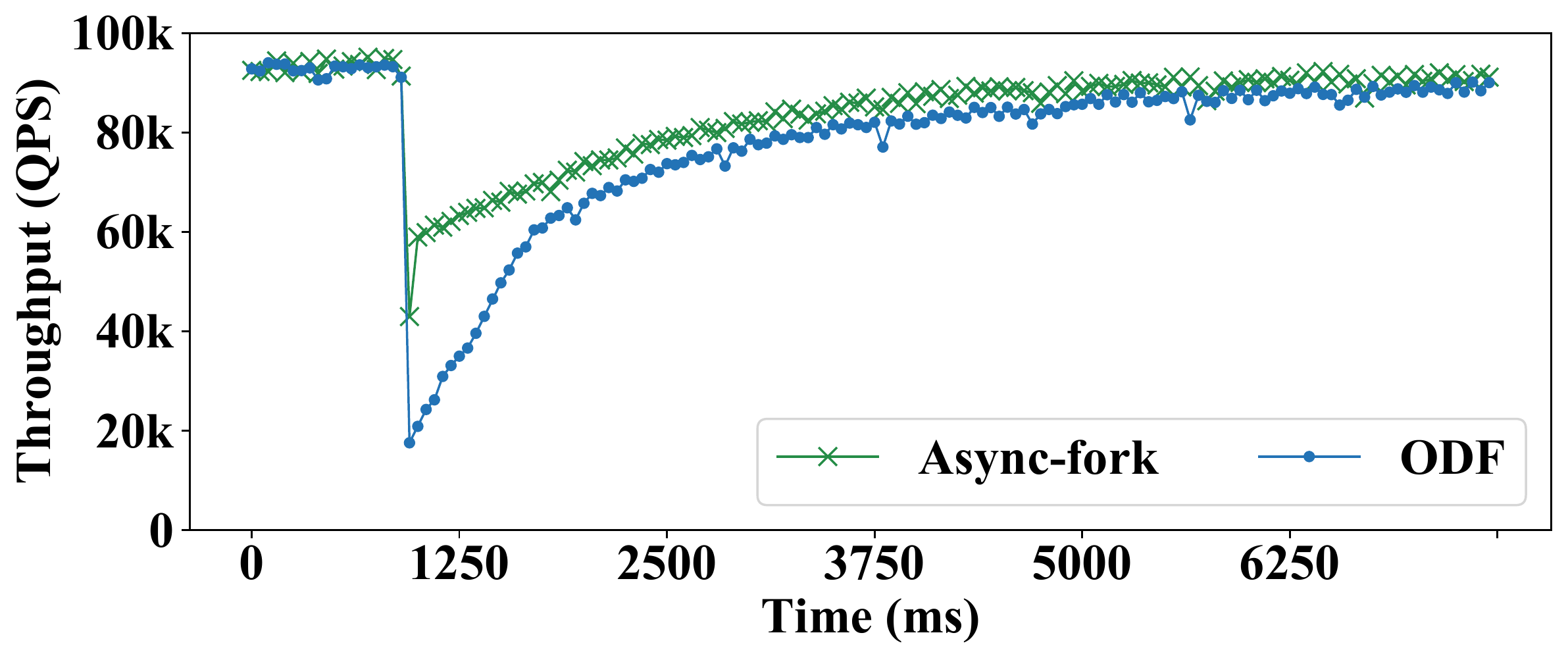}
	\caption{The changes of throughput during snapshot process in a 16GB Redis Server.}
	\label{fig:write-16gb-throughput}
%	\vspace{-2mm}
\end{figure}

\begin{figure}[t!]
	\centering
	\includegraphics[width=0.9\columnwidth]{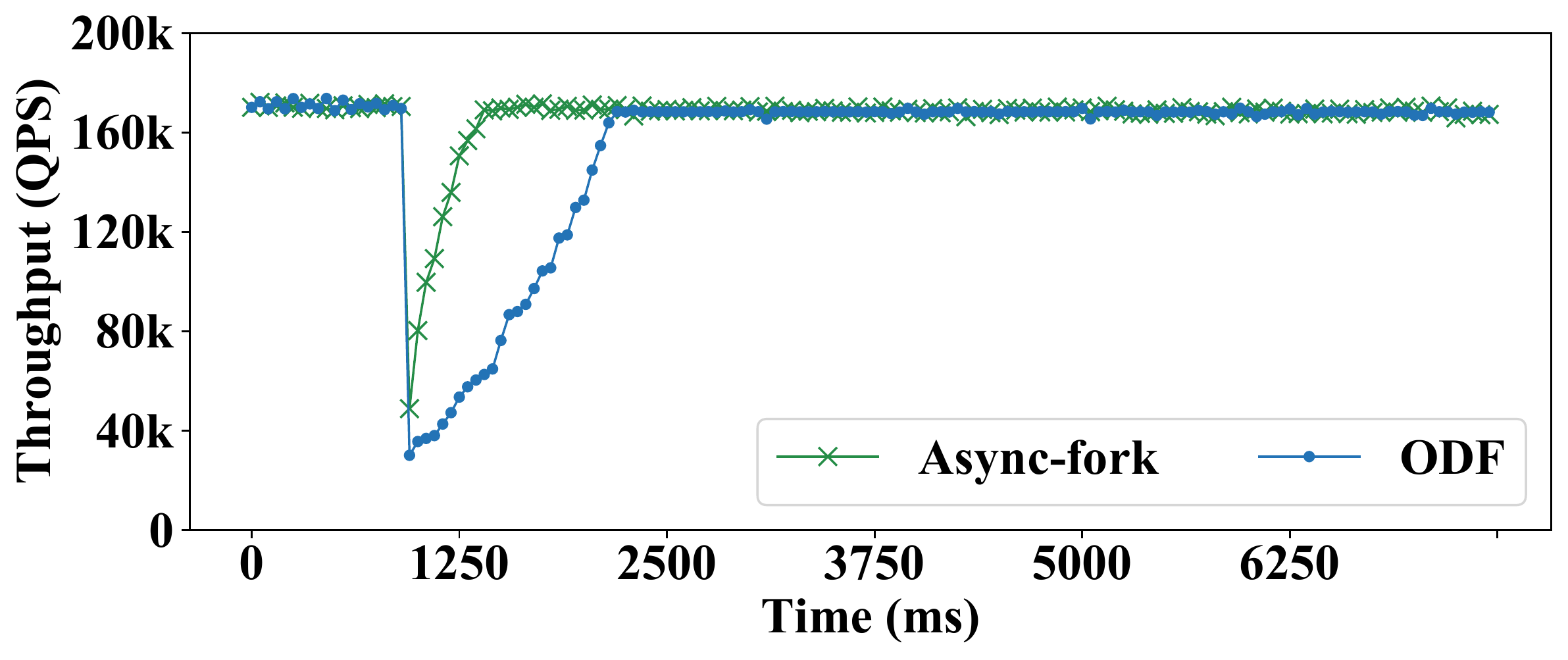}
	\caption{\pp{The changes of throughput during snapshot process in a 16GB KeyDB Server.}}
	\label{fig:write-16gb-throughput-keydb}
%	\vspace{-2mm}
\end{figure}

Figures~\ref{fig:write-16gb-throughput} and \ref{fig:write-16gb-throughput-keydb} show the processing throughput on a 16GB Redis instance and a 16GB KeyDB instance, respectively. The throughput is reported for every 50ms. %We report the number of completed queries on the server to be the processing throughput of Redis for every 50ms. 
\sun{Since KeyDB handles queries with multiple threads, we measure the throughput of KeyDB as follows: 1) each thread independently measures its throughput for every 50ms and records the value with a timestamp; and 2) for each time interval (50ms), estimate the throughput of the engine by summing up the values of all threads whose timestamps fall into the interval.}

As we can see, the throughput drops rapidly when the snapshot is taken while it increases gradually after that. 
For example, In the worst case, the throughput of Redis server drops to 17,592 queries per second (QPS) and 42,980 QPS with ODF and Async-fork respectively.
Besides, the throughput increases to the normal level much faster with Async-fork than with ODF. \sun{The throughput of KeyDB is higher than that of Redis because KeyDB uses four
threads to process queries in our experiments.}

\begin{figure}[t!]
	\centering
	\includegraphics[width=0.9\columnwidth]{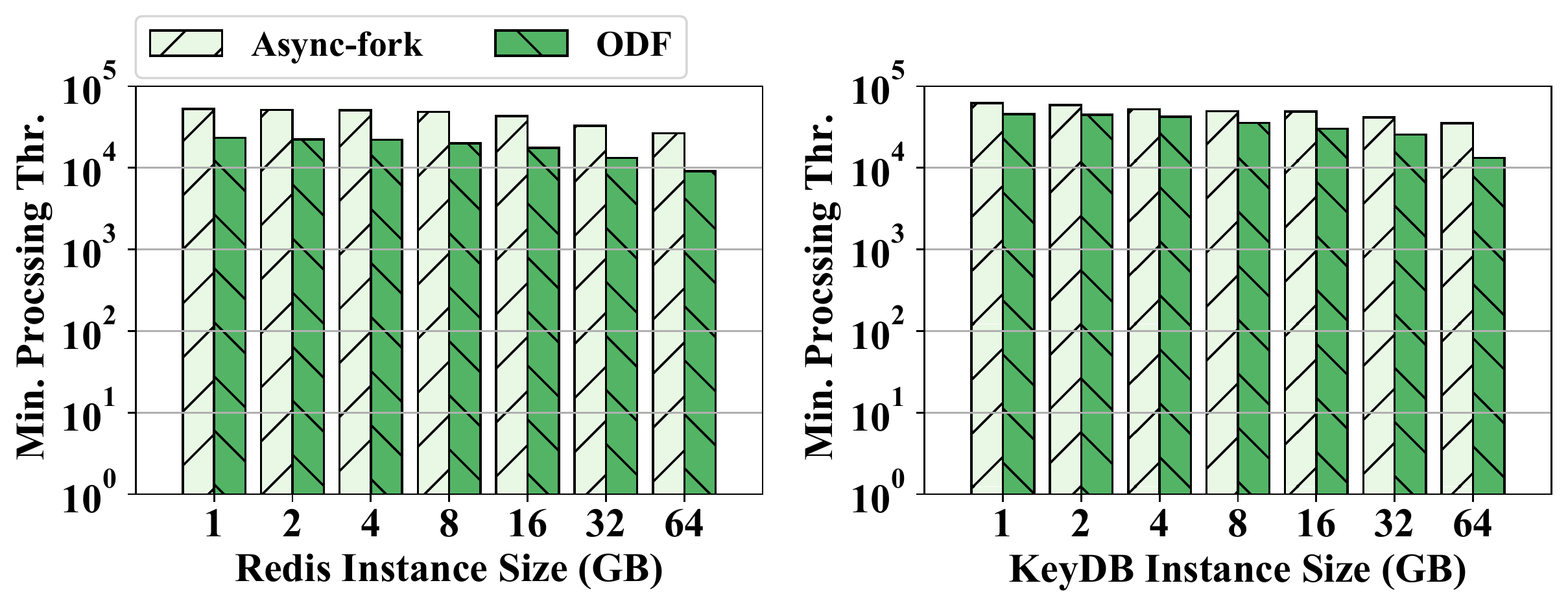}
	\caption{\pp{The minimum processing throughput during the snapshot process.}}
	\label{fig:write-low-throughput}
\end{figure}

The experiments with other instance sizes of Redis and KeyDB show similar results. 
Figure~\ref{fig:write-low-throughput} further shows the minimum processing throughput with different instance sizes of Redis and KeyDB.
Async-fork increases the minimum throughput by 2.44x on average (up to 2.9x) compared with ODF in Redis. \sun{The minimum throughput of KeyDB is increased by 1.6x on average (up to 2.69x).} Once the load exceeds the processing ability of the server, some queries queue up and suffer from a long latency.
The lower the processing throughput is, the longer the latency will be, which is consistent with the results of Section~\ref{sec:exp_write_intensive}.

\begin{figure}[t!]
	\centering
	\includegraphics[width=0.9\columnwidth]{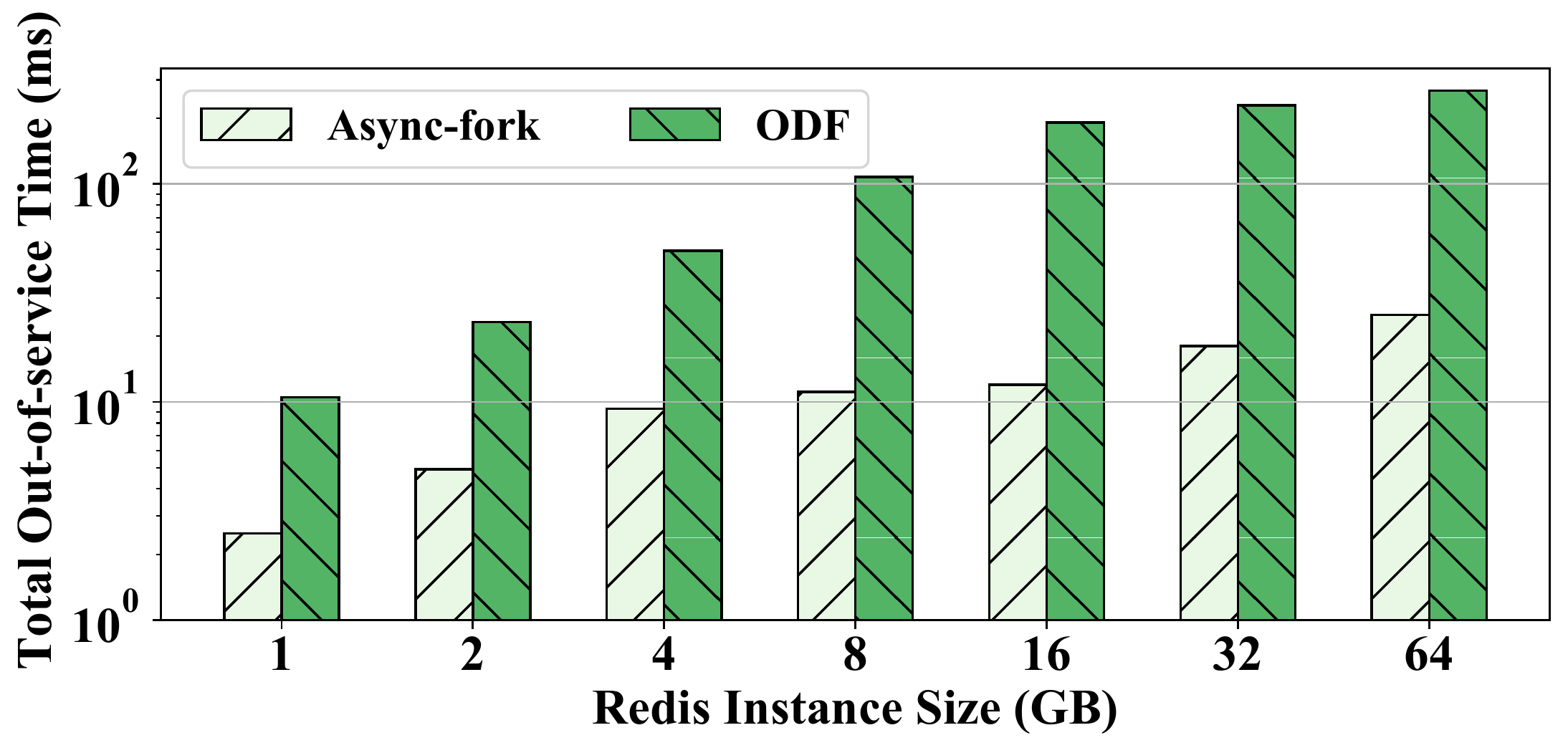}
	\caption{\pp{The total out-of-service time of Redis server.}}
	\label{fig:pmd-copy-total}
%	\vspace{-2mm}
\end{figure}

\textbf{Total Out-of-service Time.} As discussed in Section~\ref{sec:exp_write_intensive}, the parent process is out-of-service for queries during the invocation of {\it copy\_pmd\_range()}. We report the total out-of-service time (the sum of the duration of each {\it copy\_pmd\_range()} invocation), as shown in Figure~\ref{fig:pmd-copy-total}. Compared with Async-fork, the parent process is out-of-service for a longer time with ODF.

\sun{\textbf{Impact of Async-fork on Log Rewriting Queries.}
In addition to taking the snapshot, Redis can also persist data with Append Only File (AOF). Redis logs every write operation received by the server in a file, while these operations can then be replayed again at server startup to reconstruct the original dataset. However, the AOF log file gets bigger and bigger as write operations are performed. Consequently, Redis conducts the {\it log rewriting} to optimize the AOF log file~\cite{redis-persist}. Specifically, the store engine calls {\it fork} to create a child process. The child process rebuilds the AOF file into the shortest sequence of commands needed to rebuild the current dataset in memory, while the parent process continues to serve the user's queries. Due to the usage of {\it fork}, latency spikes will also be incurred during the log rewriting process.}

\sun{In the experiments, we use the Redis benchmark~\cite{redis-bench} to generate the workload and execute the BGREWRITEAOF command to trigger the log rewriting operation in Redis. We record the latency of the queries arriving during the period of conducting log rewriting (from the start of the fork operation to the end of log rewriting). We call these queries {\it log rewriting queries} for brevity.} 

\begin{figure}[t!]
	\centering
	\includegraphics[width=0.9\columnwidth]{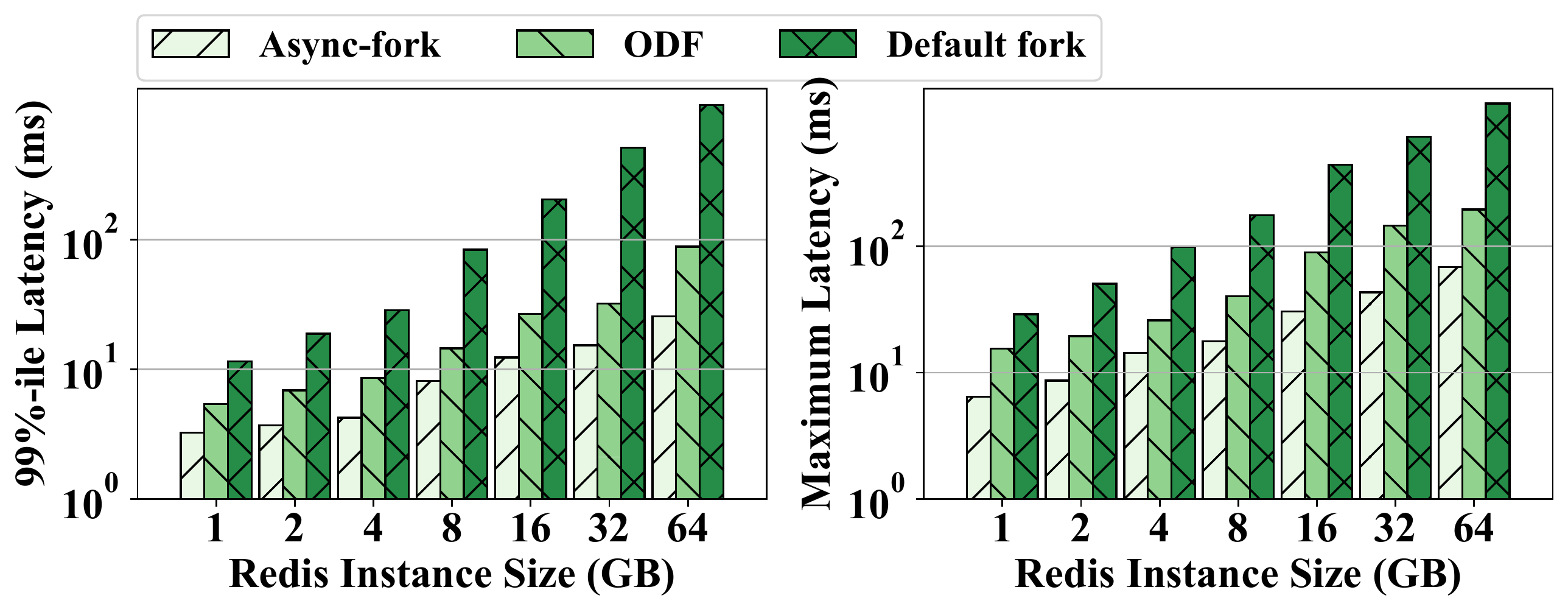}
	\caption{\sun{The 99\%-ile and maximum latency of log rewriting queries in Redis.}}
	\label{fig:aof}
\end{figure}

\sun{Figure~\ref{fig:aof} presents the 99\%-ile latency and maximum latency of log rewriting queries with different fork methods. Compared with the default {\it fork}, Async-fork reduces the 99\%-ile latency of log rewriting queries by 71.81\% (from 11.53ms to 3.25ms) on 1GB instance, 90.29\% (from 84.03ms to 8.16ms) on 8GB instance and 97.66\% (from 1093.35ms to 25.59ms) on 64GB instance. Compared with ODF, Async-fork reduces the 99\%-ile latency of log rewriting queries by 39.7\% (from 5.39ms to 3.25ms) on 1GB instance, 43.92\% (from 14.55ms to 8.16ms) on 8GB instance and 71.09\% (from 88.51ms to 25.59ms) on 64GB instance. The latency of log rewriting queries is higher than that of snapshot queries because the overall performance of Redis degrades when enabling AOF. In particular, the storage engine enabling AOF needs to synchronize the log into the disk~\cite{redis-persist} in addition to processing queries.
We observe the performance degradation of normal queries as well. When AOF is disabled, the 99\%-ile latency and maximum latency of normal queries on a 16GB instance are 0.079ms and 5.78ms, respectively. In contrast, the values increase to 1.56ms and 11.3ms when AOF is enabled. Nevertheless, Async-fork significantly reduces the latency of log rewriting queries as shown in Figure \ref{fig:aof}. The results demonstrate the effectiveness of Async-fork.}

\begin{figure}[t]
	\centering
		\includegraphics[width=0.9\columnwidth]{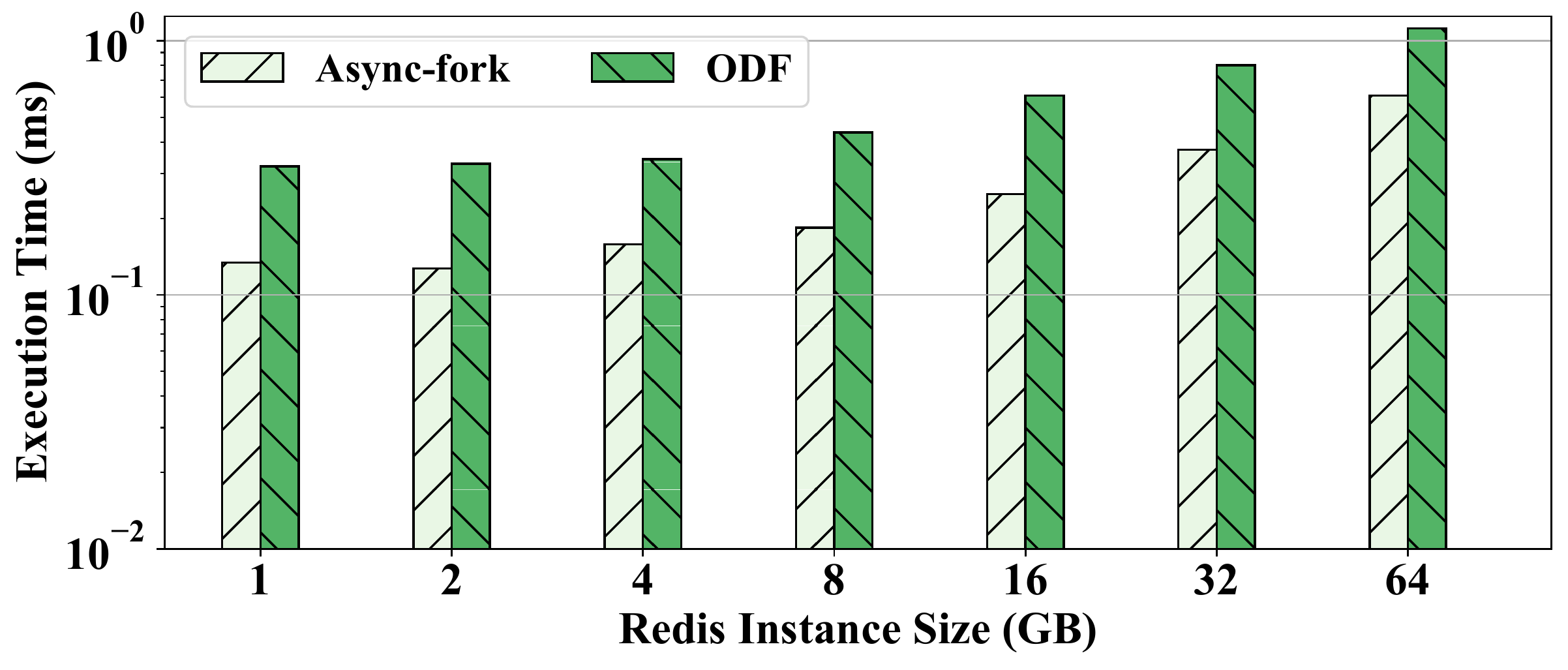}
		\caption{The execution time of Async-fork and ODF.}
		\label{fig:forktime}
\end{figure}

{\bf The Execution Time of Fork.} 
The duration of the fork operation greatly impacts the latency of the snapshot queries.
Figure~\ref{fig:forktime} shows the time of the parent process returns from calling Async-fork and ODF at different instance sizes. 
Compared with the default {\it fork} (Figure~\ref{fig:characterization_fork}), both Async-fork and ODF reduce the execution time of invoking the fork operation effectively. In the large 64GB instance, the parent process completes Async-fork in 0.61ms, and completes ODF in 1.1ms respectively. 
The time is reduced because they remove the most time-consuming phase from the default {\it fork}. We can also observe that Async-fork is slightly faster than ODF. One possible reason is that ODF introduces per-page counters to support the CoW sharing, incurring extra time cost for initialization. 	

\textbf{Tuning IMKVSes.} In our experiments, we set Redis and KeyDB to their default settings. Tuning the hyperparameters of IMKVSes may further reduce the latency of snapshot queries. In the following, we discuss the problem, while leaving it to the future studies.

By reducing the PTE modifications, the number of proactive PTE copies in Async-fork can be further reduced.
The PTE modifications can be reduced by configuring the memory allocator appropriately.
In general, by allocating a large enough virtual memory space for storing user data and meta information, the required {\it mmap} operations at runtime can be greatly reduced. In addition, it is better to retain the unused virtual memory for late use, rather than free them by calling {\it munmap}. Both the {\it mmap} and {\it munmap} operations cause PTE modifications. The default memory allocator {\it jemalloc}\footnote{\url{https://github.com/jemalloc/jemalloc}. Last accessed on 2022/11/13.} of Redis and KeyDB supports the above configuration: 1) Pre-allocating virtual memory space through {\it zmalloc/zfree} without populating it (so that no extra physical memory is consumed). 2) Configuring the built-in property `retain' to be true.

\end{document}